
\documentclass[11pt]{article}

\usepackage[letterpaper,margin=1in]{geometry}
\usepackage[T1]{fontenc}
\usepackage[utf8]{inputenc}
\usepackage{mathpazo}
\usepackage{microtype}
\usepackage{setspace}
\usepackage{booktabs}
\usepackage{threeparttable}
\usepackage{longtable}
\usepackage{tabularx}
\usepackage{array}
\usepackage{adjustbox}
\usepackage{pdflscape}
\usepackage{graphicx}
\graphicspath{{./}{coauthor_revision_2026-09-14/}{jfs/coauthor_revision_2026-09-14/}}
\usepackage{caption}
\usepackage{subcaption}
\usepackage{amsmath,amssymb}
\usepackage{enumitem}
\usepackage{placeins}
\usepackage{natbib}
\usepackage{xcolor}
\usepackage[hidelinks]{hyperref}

\hypersetup{
  pdftitle={Measuring climate credibility? Talk, emissions, and corporate debt composition},
  pdfauthor={Georgios Chortareas, Fangyuan Kou, and Alexia Ventouri}
}

\setlength{\parindent}{1.5em}
\setlength{\parskip}{0pt}
\setlength{\emergencystretch}{2em}
\onehalfspacing
\raggedbottom
\captionsetup{font=small,labelfont=bf}
\newcolumntype{L}[1]{>{\raggedright\arraybackslash}p{#1}}
\newcolumntype{C}[1]{>{\centering\arraybackslash}p{#1}}
\begin{abstract}
We examine whether a directional measure that compares emissions growth with
earlier climate discussion is associated with corporate debt composition. The
analysis links earnings-call climate language, LSEG/Refinitiv emissions,
Capital IQ debt shares, RepRisk incidents, and accounting information for
public firms over 2010--2021. The Emissions-to-Talk Index (ETI) is the central
empirical measure, but we describe it literally rather than treating it as a
general indicator of disclosure quality. In the baseline
industry-and-year fixed-effect specification, a higher ETI is associated with a
lower bank-debt share. The estimate falls by about four-fifths and is
statistically insignificant when firm and year fixed effects are used on the
same observations. It is also insignificant when the completed index is used
to predict debt composition one calendar year later. Results based on the
separate talk and emissions components, a ratio-free interaction, and complete
talk--emissions categories lead to the same broad conclusion. Oster and Frank
sensitivity calculations, alternative covariance estimates, exact year-sign
bootstrap calculations, and leave-one-year-out estimates show that the
contemporaneous  result is sensitive to specification
and inference.
We also report exploratory comparisons by competition, financial health,
opacity, and period; they do not show a
consistent within-firm pattern. DealScan tests using unique facilities likewise show no
common relation across facility incidence, spreads, maturity, security, and
covenants. The evidence supports an association in the baseline
industry-and-year specification, but not a causal claim that firms choose less bank
debt to avoid monitoring or that lenders restrict credit because of the index.


\end{abstract}

\section{Introduction}\label{sec:introduction}

Corporate climate statements now form part of the information available to
banks, bond investors, regulators and shareholders. Firms discuss transition
plans, regulation, physical hazards and new business opportunities in earnings
calls, while data providers assemble emissions records and environmental
scores. These sources may affect financing, but they do not measure the same
thing. Climate discussion may reflect a commitment, exposure to risk, an
analyst's question, or an explanation of current operations. Emissions can
reflect abatement, output, acquisitions, divestitures and reporting choices.
The relation between the two is therefore of interest, but its interpretation
requires care.

To examine this issue, we use an Emissions-to-Talk Index (ETI). The index
divides the firm's year-on-year emissions growth by the frequency of climate
language in executive sections of the previous year's earnings calls. A high
value indicates greater emissions growth per unit of earlier climate talk. It
does not directly measure the truth of a statement, a missed target, or a
manager's intent. We ask whether firms with a higher ETI use a lower share of
bank debt and consider monitoring avoidance as one possible explanation rather
than the only explanation.

Our analysis asks whether this index is related to debt composition and whether
the relation is stable across measurement and panel specifications. ETI is a directional ratio,
not a symmetric distance between a stated target and its fulfillment. The
underlying language measure counts climate-related bigrams but does not identify
whether the passage contains a promise, whether the promise concerns emissions,
or whether it was costly to make. A positive emissions change need not show
that management failed to act, and a negative change need not show deliberate
abatement. We therefore refer to ETI by its construction and use ``greenwashing''
only for externally recorded RepRisk incidents assigned to that category. This
terminology is less ambitious, but it makes the empirical question more precise:
is a firm's emissions growth relative to its earlier climate discussion
associated with its subsequent debt mix?

The question sits between two established literatures. Environmental
performance, regulation, and climate exposure are related to loan and bond
pricing \citep{chava2014,fard2020,degryse2023,kimpouget2026} and to corporate
debt composition \citep{francis2025,asimakopoulos2023,newton2024}. Work on debt choice shows that bank debt combines
private information production, monitoring and renegotiation with possible
bargaining and hold-up costs \citep{diamond1984,fama1985,diamond1991,
rajan1992,houston1996,krishnaswami1999}. These arguments do not yield a unique
sign. A borrower that dislikes scrutiny may prefer less bank debt. A bank that
views an observed discrepancy as risky may also supply less credit. Both
channels lower the observed bank-debt share. By contrast, an opaque or exposed
firm may value a bank's information production and renegotiation capacity,
which can raise its bank-debt share. An annual debt share is the outcome of
borrower and lender decisions and does not identify which side moved.

That point matters for a monitoring-avoidance interpretation. Monitoring avoidance is a
possible account of a negative coefficient, but it is not the only one. Nor
does a lower bank-debt share necessarily mean that the amount of bank borrowing
fell: a public bond issue can lower the share while bank debt is unchanged.
Corporate green bonds became more prominent during the sample period
\citep{flammer2021}. The data do not identify those bonds or
sustainability-linked loans, and they do not retain bank and bond component
amounts. We can examine the bank,
bond and residual debt shares, total debt, and accepted DealScan facilities,
but we cannot separate borrower demand from lender supply or trace every change
in the denominator.

Our data combine earnings-call climate measures, LSEG/Refinitiv Scope 1--3
emissions and environmental scores, Capital IQ debt shares, Compustat accounts,
RepRisk incidents and Sautner climate-exposure fields. The firm-year
panel contains 16,258 observations for 2,707 public firms from 2010 to 2021.
Coverage varies by variable, so each table reports its own sample. The
DealScan file contains 15,958 unique facilities. The principal loan tests use
complete calendar years 2010--2019 because the extract ends in July 2020 and
has no facilities in 2021.

We first estimate an industry-and-year fixed-effect regression and compare it
with a firm-and-year fixed-effect model on the same observations. Firm effects
absorb persistent disclosure style, production technology, financing policy
and other time-invariant firm characteristics. We then distinguish a
contemporaneous specification from a predictive one. The variable $ETI_{it}$
already combines emissions growth ending in year $t$ with talk in $t-1$. The
contemporaneous model relates it to debt composition in $t$. The predictive
model instead relates $ETI_{it}$ to debt composition in $t+1$. Finally, we examine
the separate talk and emissions components, their interaction and four complete
talk--emissions categories. These checks retain observations with zero talk and
do not divide by a small textual denominator.

We also undertake two sensitivity calculations for omitted variables.
Oster's proportional-selection statistic is reported for identical short and
full samples under each fixed-effect design \citep{oster2019}. Frank's impact
threshold for a confounding variable is reported for the baseline industry
coefficient, with an explicit qualification that Frank's calculation is based on
conventional least-squares inference \citep{frank2000,busenbark2022}. Because
the baseline has only eleven year clusters, we compare homoskedastic, firm-only,
year-only and two-way clustered standard errors, enumerate the one-way
year-cluster wild-bootstrap sign assignments, and re-estimate the model after
omitting each outcome year. None of these calculations removes endogeneity. They show
how much the conclusion depends on the stated assumptions and covariance
calculation.

The baseline result is sensitive to the fixed-effect structure. In the
contemporaneous  specification with industry and year effects,
ETI has a coefficient of $-0.0914$
($p=0.037$). On the same observations used by the firm-effect estimator, the
industry coefficient is $-0.0961$ ($p=0.016$), whereas the firm-and-year
coefficient is $-0.0178$ ($p=0.738$). The fall of about 81.5 percent is thus not
due to different observations or singleton removal. When the completed ETI is
used to predict the next year's bank-debt share, its coefficient is
statistically insignificant with either fixed-effect structure. The separate
components, continuous interaction and directional categories also yield no
clear within-firm relation with bank, bond or other-debt shares.

Inference for the baseline coefficient is equally important. Its two-way
clustered standard error is smaller than the homoskedastic, firm-clustered and
year-clustered alternatives. The associated $p$-values are 0.037, 0.169, 0.190
and 0.128, respectively. The one-way year-sign bootstrap gives $p=0.250$ when
studentized with the year-cluster covariance. The industry-effect coefficient
is negative in all eleven leave-one-year-out estimates, but is below the
5-percent threshold in only six; none of the corresponding firm-effect
estimates is below the 10-percent threshold. These results do not overturn the
industry-and-year estimate. They show that it is sensitive to the treatment of
persistent firm differences and to inference with few time clusters.

The remaining results point in the same direction. The
competition, financial-health, information-opacity and post-Paris patterns are
visible mainly in industry-effect specifications. Most interactions are small
and statistically insignificant with firm effects; the few raw exceptions are
not persuasive after allowing for the number of related tests. RepRisk total
and environmental incident counts have modest negative associations with bank
share in some firm-effect models, but greenwashing-specific incidents do not.
Those results concern external negative news and cannot be used as a second
measure of ETI. DealScan regressions show no common pattern across
facility appearance, amounts, spreads, maturity, security and covenants.

Our contribution is a detailed examination of a talk--emissions construction
in a debt-choice setting. We compare
industry-and-year estimates with same-sample
firm-and-year estimates and show that the negative ETI association attenuates
when persistent firm differences are absorbed. We also distinguish climate-language salience,
operational emissions, topical exposure, vendor scores and external incidents
instead of treating them as synonyms. This comparison complements studies that
relate climate exposure, ESG ratings and adverse news to debt structure
\citep{francis2025,asimakopoulos2023,newton2024}. It does not establish a new
causal mechanism. The interpretation depends on the construction of the ratio, and a debt share
alone cannot distinguish borrower from lender behavior.

The paper proceeds as follows. Section~\ref{sec:framework} discusses the
measurement and debt-choice arguments. Section~\ref{sec:data} describes the
data and construction of ETI and the comparison measures.
Section~\ref{sec:design} presents the empirical design and inference.
Section~\ref{sec:results} reports the measurement evidence and central debt-
composition results. Section~\ref{sec:heterogeneity} presents the
heterogeneity analyses. Section~\ref{sec:additional} reports external
incidents, alternative measures and DealScan tests. The final sections discuss
interpretation, limitations and conclusions.

\section{Climate information and debt choice}\label{sec:framework}

\subsection{Words, environmental outcomes, and external events}

We distinguish four measures that are observable in the data. The first is \emph{climate-language
salience}: how much of an earnings-call segment uses climate-related language.
The Sautner dictionary was designed to measure participants' attention to
perceived firm-level climate exposure \citep{sautner2023}. It can capture
discussion of an opportunity,
physical exposure, or regulatory risk.  This breadth suits exposure measurement but prevents a direct interpretation
as favorable disclosure.
Even prepared remarks are not necessarily voluntary claims about future
emissions; they may address a risk that investors already regard as material.

The second object is an \emph{emissions outcome}.  Scope 1, Scope 2, and Scope 3
emissions refer to different operational boundaries and are subject to unequal
coverage.  Total emissions offer broader coverage but mix sources and may
include estimates.  Levels primarily reflect scale and production technology;
intensity partly adjusts for scale; changes capture movement relative to a
firm's recent state.  None maps one-for-one into managerial effort.  For
example, a divestiture can lower reported emissions without abatement, and a
growing firm can increase total emissions while lowering intensity.

The third is a \emph{talk--emissions configuration}. It joins two
observables without claiming that either measures intent. One simple,
directly observed configuration is high presentation talk accompanied by rising emissions. Its
meaning is directional and easily
stated.  It can flag an observation for further investigation, but it is not a
distance between a specific promise and its fulfillment.  A continuous
interaction has an equally limited interpretation: it asks whether the
association of one standardized component depends on the other.

The fourth is an \emph{externally recorded controversy}. RepRisk
collects event-based negative news, including environmental incidents and a
greenwashing category \citep{reprisk2022}. Such records originate outside the firm's own call and
are therefore recorded independently of it. They still depend on media
coverage, the provider's classification rules, and issuer matching, and should not be described
as legal findings.  Total ESG events also combine environmental, social, and
governance content.  An association for total incidents is not evidence of a
climate-specific effect; an association for environmental incidents need not
be greenwashing-specific.

These distinctions matter for the signaling argument. Cheap talk can be informative
in equilibrium, but a costly-signaling interpretation requires the content of
a commitment, a cost that differs by type, and conditions limiting imitation.
The frequency variables do not identify those elements.  Emissions
abatement and verified green investment may be costly; uttering a climate
bigram is not demonstrated to be costly.  Accordingly, we do not formulate a
directional hypothesis in which more climate talk automatically signals better
environmental quality.  We ask instead whether the segments predict subsequent
emissions or external events after conditioning on prior state.

Our data distinguish presentation- and answer-section frequencies. If the
underlying assignments are accurate, their information processes differ:
prepared remarks are selected before the call,
whereas answers respond to questions and may reflect contemporaneous information
demand. The raw transcripts and parsing files are unavailable, so we cannot
independently verify those assignments. A joint regression asks whether each
segment measure contains incremental information conditional on the
other. We treat any difference as an exploratory, descriptive segment-specific association,
not evidence of motive, agency, or causality.

\subsection{Competing implications for bank debt}

Classic debt-choice theory offers no unconditional sign for the relation
between climate information and bank debt. Banks economize on duplicated
monitoring and produce private information \citep{diamond1984,fama1985}.
Private-debt placement and syndicated-loan structure respond to information
asymmetry and monitoring needs \citep{krishnaswami1999,sufi2007}. The
value of that information can be greatest when public signals are incomplete.
Private debt can also facilitate renegotiation when uncertainty is high.
These forces imply that a firm with difficult-to-evaluate climate exposure may
use more bank debt.

Informed lending has costs as well.  A bank's informational advantage can
affect bargaining and create hold-up concerns \citep{rajan1992,houston1996}.
Managers may dislike investigation of statements that appear inconsistent with
outcomes.  If the perceived cost of scrutiny exceeds the information and
renegotiation benefits, borrower demand for bank debt may fall.  This is a
monitoring-avoidance account.  Its sign is
conditional on an unobserved tradeoff and is not implied by delegated-monitoring
theory alone.

Lender screening can generate the same observed sign for a different reason.
If a bank interprets an external controversy or an apparent talk--emissions
discrepancy as transition, governance, or reputation risk, it may restrict
credit or change contract terms.  Studies with lender identities and lender
environmental commitments show what is required to examine this channel
\citep{degryse2023,hasan2026,kacperczykpeydro2026}.  Borrower demand and lender supply jointly
determine an accepted facility.  Without rejected applications, lender
attributes, or plausibly exogenous changes in lender policy, the two cannot be
separated.

The loan data permit a limited check on these explanations. A screening account
would ordinarily predict some combination of fewer recorded
facilities, smaller reported amounts, higher spreads, shorter maturity, more
security, or more covenants.  A monitoring-value account could instead predict
more bank borrowing and contract structures that facilitate oversight.  A
debt-share result unaccompanied by any loan-incidence or contract-term pattern
provides only weak evidence about either explanation. Evidence based on accepted
facilities remains selected, but considering several terms together is more
informative than a bank share alone.

Debt composition also has an accounting dimension.  Bank and bond shares are
ratios to total debt, and firms hold other private or non-bank debt.  The debt
mix varies substantially within firms and across credit quality
\citep{denismihov2003,rauhsufi2010}.  An increase in bond issuance can reduce
the bank share without reducing bank debt.  Green bonds add an instrument that
could matter for this substitution \citep{flammer2021}.  We therefore examine
bank, bond, and residual other-debt shares together, and separately check total
debt and debt-to-assets.  We cannot identify green bonds or
sustainability-linked loans in the available data, so we avoid attributing a
share movement to those instruments.

These arguments guide the empirical tests without determining their outcome.
Our main question is whether the ETI is associated with debt composition
once persistent firm characteristics are absorbed and timing is made explicit.
The analysis has three parts: whether such a relation is present in the panel
estimates; whether it arises from the combined ratio or from its talk and
emissions components; and whether the debt-share pattern is
accompanied by loan-incidence or contract-term evidence. Finding a negative
coefficient in the panel estimates would be consistent with both monitoring
avoidance and lender screening. Finding no statistically detectable association would not show that
banks disregard climate information; it would show only that this index has no
detectable incremental relation with annual debt composition in the stated
model.

\subsection{Relation to climate finance and debt-choice evidence}

The closest climate-finance studies examine several distinct margins.
\citet{chava2014} links environmental externalities to the cost of capital,
while \citet{fard2020} studies environmental regulation and bank-loan costs.
\citet{degryse2023} uses lender and borrower environmental characteristics to
study loan contracting. \citet{kimpouget2026} separates primary from secondary
corporate bond markets and finds that the emissions relation differs across
those settings. These papers make clear why an annual bank-debt share should
not be described as ``access'' to public debt: it is neither an issuance
decision nor a price.

Recent work is closer to the debt-mix outcome. \citet{francis2025} relates
climate risk to debt structure, \citet{asimakopoulos2023} considers ESG ratings,
and \citet{newton2024} examines external ESG reputation risk. Their explanatory
variables do not have the same provenance or interpretation as ETI. Exposure
can increase the value of relationship lending and contractual flexibility;
negative news can affect perceived risk; an ESG rating can affect investors'
screening; and a talk--emissions ratio combines a firm's words with an
operational outcome. A finding for one is not a benchmark sign that every other
measure should reproduce.

Research on greenwashing raises a separate construct issue. A text-based or
machine-learning classification can be designed to identify misleading claims,
as in \citet{wang2024}'s study of Chinese firms. Our call-frequency
field was not trained for that task. It comes from the Sautner climate-exposure
dictionary, whose purpose is to locate discussion of climate opportunities,
physical exposure and regulation \citep{sautner2023}. We therefore compare ETI
with RepRisk's specifically coded greenwashing incidents but do not treat
agreement as automatic. RepRisk records negative news and is itself affected by
coverage and classification. We therefore use RepRisk as an external
comparison, not as a definitive classification.

Our empirical contribution is to bring these distinctions to the debt-choice
question. The baseline
industry-and-year specification produces a negative association. The other
tests ask whether that
relation is present within firms, predicts a later debt mix, appears in ratio-free versions,
or is accompanied by loan outcomes. This is a narrower contribution than
claiming that ETI identifies greenwashing or monitoring avoidance. The available data support this narrower question.

\section{Data and measurement}\label{sec:data}

\subsection{Firm-year and facility samples}

Our master panel is indexed by Compustat \texttt{gvkey} and calendar
year.  It contains 16,258 unique firm-year observations for 2,707 firms from
2010 through 2021. Table~\ref{tab:jfs_sample} reports the sample flow,
coverage, and descriptive statistics. Coverage differs across variables, so
each regression table identifies its estimation sample, and either its notes
or the sample-flow table reports the relevant firm and outcome-year cluster
counts; we do not invoke a single ``final sample.'' The
panel is unbalanced and contains 141 gaps.  Exact-calendar changes and lags are
missing when year $t-1$ is not observed.

The panel brings together precomputed earnings-call variables;
LSEG/Refinitiv emissions, ESG, and Emissions-score fields; Capital IQ debt
shares; Compustat accounting variables; RepRisk event counts; and Sautner
climate-exposure variables. Our data begin with the annual call measures and
vendor fields; they do not include the transcript-extraction files or the full
sequence of initial merges. This matters for
ETI, whose values can be used in the regressions but whose
upstream text scaling, floor, and zero treatment cannot be regenerated exactly.

The DealScan file contains 15,958 unique facility identifiers for 2,631 firms
from 2010 through 2020. We use one record per facility throughout. Of these
facilities, 15,325 occur in the ten complete
calendar years 2010--2019, of which 6,677 match the master panel on
\texttt{gvkey} and year before regression-variable requirements. Across the
full 2010--partial-2020 file, 6,866 facilities match. A failed match is not
coded as an absence of borrowing.  Facility coverage in 2020 stops on July 13,
and no 2021 facilities are present; those years are excluded from loan-incidence
and principal contract tests.

\subsection{Earnings-call climate language}

The earnings-call measures use climate-related bigrams based on
\citet{sautner2023}.  Let $C$ denote the climate dictionary and $B^s_{ict}$ the
number of eligible bigrams in segment $s$ of call $c$ for firm $i$ in year $t$.
The conceptual segment frequency is
\begin{equation}
 Talk^s_{ict}=\frac{1}{B^s_{ict}}
 \sum_{b=1}^{B^s_{ict}}\mathbf{1}\{b\in C\},
 \qquad s\in\{P,A\},
 \label{eq:talk}
\end{equation}
where $P$ denotes the management presentation and $A$ management answers. The
data contain annual, precomputed segment frequencies. Because they do not
include the upstream transcripts and aggregation program, we describe the
available fields rather than claiming to re-create speaker parsing or
call-to-year weighting.

Presentation and answer frequencies are nonnegative, right-skewed, and have
many observed zeros.  The exploratory segment analysis therefore applies the
transformation
\begin{equation}
 \widetilde{Talk}^{s}_{it}=
 z\!\left[\log\{1+10{,}000\,Talk^{s}_{it}\}\right],
 \label{eq:talktransform}
\end{equation}
where $z[\cdot]$ standardizes over the observed firm-year sample.  The logarithm
is defined at zero and preserves ordering; standardization then centers and
scales the transformed measure. Both segments enter the segment-comparison regression
jointly.  For the debt analysis, we standardize presentation talk within
two-digit SIC industry-year cells, so the variable compares a firm's climate
salience with contemporaneous firms in the same broad industry.  Cells with
fewer than three observations are excluded from standardization.

We call these fields climate-language salience. The presentation field is more
firm-directed than the answer field, but it still may
describe adverse exposure rather than environmental commitment. The Sautner overall, opportunity,
regulatory, and physical exposure measures are comparison variables.  A strong
correlation with them is expected because the measures share a dictionary and
transcript source; it shows common climate-language content, not independent
evidence of green quality.

\subsection{Emissions and climate-information constructions}

Let $E_{it}$ denote the available LSEG/Refinitiv total Scope 1--3 field.  We
examine $\log E_{it}$, the exact-calendar change
\begin{equation}
 \Delta\log E_{it}=\log E_{it}-\log E_{i,t-1},
 \label{eq:emissionschange}
\end{equation}
and log emissions intensity, defined using revenue.  Total emissions provide
much broader coverage than the scope-specific fields. The data
do not identify, observation by observation, whether the total is reported,
estimated, or carried forward, and they do not preserve historical data
vintages.  Total emissions are exactly unchanged from the preceding year in
3,929 firm-years. We keep these observations but do not assume that equality
means no change in underlying operations.

We define the Emissions-to-Talk Index (ETI) as percentage emissions growth
divided by lagged presentation-plus-management-answer talk:
\begin{equation}
 ETI_{it}=\frac{(E_{it}-E_{i,t-1})/E_{i,t-1}}{Talk^{PA}_{i,t-1}}.
 \label{eq:originaleti}
\end{equation}
 We use the  ETI in the baseline regressions and assess its timing and
distributional sensitivity, winsorizing it at the 1st and 99th percentiles in
the distributional checks.

The  data contain two additional ratios used only for comparison: emissions growth divided
by lagged presentation talk and the change in log emissions divided by the
change in presentation talk. Appendix Table~\ref{tab:jfs_ratio_distribution}
summarizes  these ratios. We also  form exact-calendar versions from
 their components, setting a zero denominator to missing, and compare
them before and after fixed 1st/99th-percentile winsorization in Appendix
Table~\ref{tab:jfs_ratio_tail_diagnostics}. These ratios are not central to the
interpretation because a small denominator can assign enormous leverage
to an observation for reasons unrelated to the economic discrepancy of
interest.

The principal ratio-free checks avoid division. Within two-digit SIC
industry-year cells, let
\begin{equation}
 T_{it}=z_{st}(Talk^P_{it}),\qquad
 D_{it}=z_{st}(\Delta\log E_{it}),\qquad
 I_{it}=z(T_{it}D_{it}),
\end{equation}
where the outer $z(\cdot)$ restandardizes the product over the panel.
The continuous model enters $T_{it}$, $D_{it}$, and $I_{it}$ jointly.
The main ratio-free and common-sample models  scale continuous
information measures to a one-standard-deviation change and use information
dated $t-1$. ETI, the  comparison ratios,  the moderator models, and
the  DealScan specifications retain the units and timing
stated in their table notes. We
also form four complete quadrants according
to whether presentation talk is above versus at or below its industry-year
median and whether emissions rise. Low talk with non-rising emissions is the reference category;
the other indicators are high talk/non-rising emissions, low talk/rising
emissions, and high talk/rising emissions. The last category is a simple,
observable configuration, not a classification of managerial intent.

The main ratio-free, segment, and complete-year DealScan analyses use information
from firm-year $t$ to predict an outcome in $t+1$. Equivalently, those
regressions place an exact one-year calendar lag on the right-hand side. The
 ETI and the ratio comparisons retain their contemporaneous
timing and are labeled accordingly. This matters for ETI because the index
already uses emissions growth ending in its stated
year and talk from the preceding year.  Using the index contemporaneously with
debt is not the same as asking whether the completed
index predicts later debt structure.

\subsection{Debt, controversy, score, and loan outcomes}

Capital IQ supplies bank debt and public bond debt as shares of total debt.  We
define other-debt share as one minus the two reported shares. The three shares
sum to one up to storage precision.  Examining them jointly helps identify
whether an estimate is bank-specific, but the data do not contain the
component dollar amounts needed to decompose a change in ratios.  We use
Compustat long- and short-term debt to construct total debt and debt-to-assets
only to check the role of the denominator; we do not infer bank or bond amounts by
multiplying shares by Compustat debt.

RepRisk fields count externally recorded negative-news incidents
\citep{reprisk2022}. The main
criteria are an indicator for at least one greenwashing incident,
$\log(1+n)$ greenwashing incidents, $\log(1+n)$ environmental incidents, and
$\log(1+n)$ total ESG incidents.  Continuous incident counts are standardized
when used as financing predictors. In comparisons with later external outcomes,
those outcomes are dated one year after talk, and the specifications include
their exact calendar lags. For financing, controversies are dated one year
before the debt or loan outcome.

The simultaneous information comparison also includes standardized Sautner
opportunity, regulatory, and physical exposure, and standardized LSEG ESG and
Emissions scores.  These measures are deliberately not pooled into a composite.
Their source and content differ: the topical variables are call-based exposure,
the LSEG fields are vendor scores, and RepRisk counts external negative news.
The joint regression asks whether any measure adds information about debt
composition after conditioning on all the others.

At the firm-year level, DealScan outcomes are an indicator for at least one
matched facility, $\log(1+\text{facility count})$,
$\log(1+\text{reported facility amount})$, average all-in-drawn spread, and
average maturity.  ``No matched facility'' means no facility observed in the
DealScan extract for that firm-year; it does not mean an application was
rejected.  At the facility level, outcomes are log all-in-drawn spread, log
maturity, an explicitly reported secured indicator, the number of financial
covenants, and revolver status.  The source units of reported facility amount
are not treated as verified U.S. dollars in the all-currency sample.  We report
a U.S.-dollar-only sensitivity analysis separately.  In the principal loan analyses, unknown security status,
which occurs for 5,518 facilities in the facility file, remains missing rather
than being coded unsecured.

\subsection{Controls}

Firm-year controls are log assets, total debt divided by assets, net property,
plant and equipment divided by assets, net income divided by assets, Tobin's
$Q$, and the winsorized Altman $Z$-score \citep{altman1968}. Financing and
controversy models also include the prior emissions state as appropriate. The
emissions tests in Table~\ref{tab:jfs_talk_emissions}  build the control set
from  Compustat components. Other
predictive models  use exact-calendar lags and  build debt-to-assets as
$(dltt+dlc)/at$ and the other accounting ratios from Compustat components.
Some specifications use the precomputed control fields. Their leverage
definition differs from debt/assets, and several fields align with later
inputs. The predictive specifications instead use controls reconstructed at
exact calendar lags.

Facility regressions add the relevant contract controls. Spread models include
log reported facility amount, log maturity, and security status; maturity models
include amount and security; security models include amount and maturity; and
covenant and revolver models include amount, maturity, and security. We also add
pooled loan-type, deal-purpose, and currency categories. Source
categories represented by fewer than 20 facilities are pooled to avoid cells
that cannot support the fixed-effect comparison.

\section{Empirical design}\label{sec:design}

\subsection{Segment-level comparisons with external outcomes}

We first examine whether prepared and answer-segment talk contain different
information about subsequent outcomes.  For outcome $V_{it}$, the primary specification is
\begin{equation}
 V_{it}=\beta_P\widetilde{Talk}^{P}_{i,t-1}
       +\beta_A\widetilde{Talk}^{A}_{i,t-1}
       +\rho V_{i,t-1}+\theta'X_{i,t-1}
       +\mu_i+\tau_t+\varepsilon_{it}.
 \label{eq:segment}
\end{equation}
The outcomes are emissions change, LSEG scores, greenwashing incidence and
count, environmental incidents, and incidents classified as misleading
communication. The lagged dependent variable is included in the persistence
specification reported in the main table; for emissions change it reduces the
sample because two consecutive calendar changes are required.  The short panel
means this dynamic fixed-effect model should be read only as a predictive
exercise, not as a structural dynamic model. We also estimate the equation
without $V_{i,t-1}$ and use industry effects in place of firm effects on the
identical sample.

\subsection{Debt-composition models}

For debt outcome $Y_{it}$, the central model is
\begin{equation}
 Y_{it}=\alpha+\beta ETI_{i,t-h}+\phi\log E_{i,t-1}
       +\theta'X_{i,t-1}
       +\mu_i+\tau_t+u_{it}.
 \label{eq:debt}
\end{equation}
We use two values of $h$. With $h=0$, the emissions change embedded in
$ETI_{it}$ and debt share $Y_{it}$ refer to the same year, while the talk
denominator refers to $t-1$. With $h=1$, the completed index at $t-1$ is
related to debt share at $t$. The latter is the predictive specification. We
report both because they answer different timing questions.

We estimate Equation~\eqref{eq:debt} for bank, bond, and other-debt shares.
Firm effects absorb time-invariant business models, average disclosure styles,
production technologies, industry affiliation, and persistent debt policies.
Year effects absorb common shocks. Under $h=1$, $\beta$ is the conditional
association between a change in a firm's completed ETI and its debt share one
year later. Under $h=0$, it is a contemporaneous conditional association and
does not establish temporal ordering.

The same-sample comparison replaces $\mu_i$ with two-digit SIC fixed effects and
retains year effects. Its sample is exactly the sample used by the firm-effect
model after recursive singleton removal, so differences between the two
coefficients cannot be attributed to missing observations. The industry model
combines between- and within-firm variation; it is not a second primary result.

The ratio-free models are measurement checks on the same question. For climate-
information vector $G_{i,t-1}$, we estimate
\begin{equation}
 Y_{it}=\alpha+\beta'G_{i,t-1}+\theta'X_{i,t-1}
       +\mu_i+\tau_t+u_{it}.
 \label{eq:debt_ratiofree}
\end{equation}
In the continuous specification, $G$ contains presentation talk, emissions
change and their interaction. In the categorical specification, it contains
the high-talk/non-rising, low-talk/rising and high-talk/rising cells, with low
talk/non-rising emissions omitted. These models ask whether the conclusion
depends on division by talk. They are not alternative definitions of ETI chosen
to produce a particular sign.

Debt shares are observed proportions, not censored realizations of a latent
variable, so we use linear fixed-effect models. Binary outcomes use
linear-probability models as descriptive checks, and counts use $\log(1+n)$. We emphasize
direction, stability, and family-wise inference for
these bounded and sparse outcomes.

The full comparison expands $G$ to include presentation talk, emissions change,
their interaction, RepRisk environmental incidents, a RepRisk greenwashing
indicator, the three Sautner topical exposures, and the two LSEG scores.  All
ten information variables enter simultaneously.  Continuous measures are standardized. We ask whether any source has an
additional association with debt composition after conditioning on the others
in a common sample.

\subsection{Controversy, total-debt, and DealScan tests}

The controversy analysis applies Equation~\eqref{eq:debt} first to
standardized environmental incidents and the greenwashing indicator, then to
total incidents alone, and finally to environmental and total incidents
together. The conditional model asks whether the relation is specific to
environmental incidents. If the environmental coefficient loses magnitude and
statistical significance after total incidents are included, it does not
provide distinct evidence of an environmental association. We also estimate a
distributed model with leads and lags and descriptive event-time coefficients
around the first observed positive environmental count. The event-time model
is explicitly descriptive.  Incident recording is not a staggered treatment,
firms can have repeated events, and the pre-event coefficients do not support a
causal design.

Total-debt checks replace the share outcome with log total debt or total
debt-to-assets and add the lagged dependent variable.  They ask whether a share
association is accompanied by a change in overall borrowing.  They cannot
recover bank-debt or bond-debt amounts.

The DealScan models retain the timing and fixed-effect structure of Equation
~\eqref{eq:debt}. Firm-year models examine appearance and volume in the
DealScan extract. Facility models condition on the observed contract and
therefore estimate associations among accepted facilities. The complete-year
analysis clusters by firm and outcome year. The supplementary contemporaneous ETI models for non-price facility terms
in Table~\ref{tab:jfs_dealscan_terms} include partial 2020 and use firm
clusters only.
Because multiple facilities may belong to the same firm-year, the firm cluster
accounts for repeated borrower observations, while the year cluster in the
complete-year analysis accounts for common credit conditions.

\subsection{Inference and multiplicity}

Principal firm-year standard errors are clustered two ways, by firm and year
\citep{cameron2011}. Reported
$p$-values and confidence intervals use a $t$ reference distribution with
degrees of freedom based on the smaller cluster dimension.  Depending on the
sample, there are only eight to eleven year clusters.  We therefore report the
cluster counts prominently and treat conventional threshold crossings as
suggestive unless they are stable across these checks.

We also use alternative covariance estimators, leave-one-year-out ranges, and
an exhaustively enumerated, null-imposed one-way year-cluster
wild-bootstrap-$t$ for six coefficients selected in advance
\citep{cameron2008,roodman2019}. The last calculation
removes Monte Carlo error but omits within-firm dependence and therefore does
not replace the primary estimator.

Holm $p$-values control family-wise error within each declared specification
family \citep{holm1979}; the relevant table notes list the terms or outcomes
included in that family. Broader section-wide and global adjustments are
reported where relevant. Raw values show where small $p$-values arise, while
the adjusted values guard against placing weight on isolated results from a
large set of related outcomes. We apply the stated timing,
fixed-effect, and multiplicity checks across each declared family and do not
interpret isolated raw threshold crossings as primary evidence.

If the multiway covariance is indefinite, we use the eigenvalue correction
described by \citet{cameron2011}: negative eigenvalues are set to zero and the
model is flagged. This issue is concentrated in expanded DealScan category
grids; the primary ratio-free and separate RepRisk debt-share models do not
require the correction.
Table~\ref{tab:jfs_inference} reports the clustering, leave-one-year-out, and
wild-bootstrap checks; the supplement reports the wider multiplicity
calculations.

We report two further sensitivity calculations for the central ETI coefficient.
Following \citet{oster2019}, short and full regressions use identical samples
and the same fixed-effect transformation. One comparison adds lagged emissions
and all six financial controls to an ETI-only regression; a second holds lagged
emissions in both models and adds the six financial controls. We report signed
values of the proportional-selection parameter for $R_{\max}=1.3R^2_{full}$
and $R_{\max}=1$, because the result is sensitive to that choice. Following
\citet{frank2000}, we also report the impact threshold for an omitted
confounder and the corresponding equal partial correlations. The conventional
formula does not directly accommodate two-way clustered inference, so this is
reported as a sensitivity calculation rather than a formal clustered test.

We do not use the state-level ESG instrument. State ESG conditions
can affect environmental policy, litigation, investors, banking conditions and
debt markets directly, so its exclusion restriction is not credible. We also
do not use extreme-quartile propensity-score matching as evidence
about unobserved selection. Removing those analyses is preferable to retaining
them with claims they cannot support.

\section{Results}\label{sec:results}

\subsection{Sample coverage and measurement checks}

Table~\ref{tab:jfs_sample} documents the movement from the 16,258-row master
panel to the principal estimation samples and summarizes the main variables.
Debt shares are observed for 15,912
firm-years.  Exact lags, complete controls, and fixed-effect singleton removal
leave 8,219 observations for the continuous ratio-free debt model and 8,242 for
the quadrant model.  The RepRisk debt model uses 10,373 observations.  The
simultaneous ten-measure comparison uses 8,061 observations.  Facility-level
spread tests are smaller: the continuous talk--emissions specification uses
1,946 facilities from 425 firms and eight outcome years.  These differences
make table-specific samples and cluster counts integral to interpretation.

ETI is available for 13,410 observations. It has a mean of 0.0030,
a standard deviation of 0.0439, and an observed range from $-1.1368$ to 1.5409.
Appendix Table~\ref{tab:jfs_ratio_distribution} reports its full empirical
distribution and central mass, and
Figure~\ref{fig:submitted_eti_distribution} plots the distribution. The ETI is much less dispersed than ratios computed directly from the
available emissions and talk components. The source files do not establish why.

The  presentation-only comparison ratio has 9,519 nonmissing values, a standard
deviation of approximately 1,604, and a range from approximately $-8{,}117$ to
$69{,}705$. The  change-on-change ratio is more extreme: among 11,394 values its
standard deviation is approximately 382,359 and its range extends from about
$-40.24$ million to $2.62$ million. Appendix
Table~\ref{tab:jfs_ratio_distribution} and
Figure~\ref{fig:all_ratio_distributions} show that these
tails are not minor departures from normality.  They follow mechanically when
presentation talk or its change approaches zero.

The distributional problem has empirical consequences. In the
 exact-lag comparisons, the untrimmed presentation-only ratio has no statistically detectable association with bank
share in the firm-effect model,
whereas its 1st/99th-percentile winsorized version has a positive coefficient.
That direction is opposite to the monitoring-avoidance prediction.
Other ratio estimates also change across tail-treatment rules and debt categories
(Appendix Table~\ref{tab:jfs_ratio_tail_diagnostics}).
For this reason, statistical significance after a particular winsorization is
not treated as confirmation.  All central conclusions are based on components
and quadrants that do not divide by talk.

The sample correlations in Appendix Table~\ref{tab:jfs_correlations}
support the same measurement distinction.  Presentation talk is highly
correlated with the Sautner overall exposure measure (Pearson correlation
0.966), as expected from shared dictionary content.  That number shows that the
field captures climate-related language.  It cannot independently establish
that the language is favorable, costly, or misleading.  Correlations among
RepRisk, LSEG, emissions, and debt fields are substantially less mechanical,
which motivates the joint models rather than substituting one label for another.

Appendix Table~\ref{tab:submitted_correlations} reports correlations among the
ETI, log total emissions, presentation-plus-management-answer TalkFreq, the
LSEG ESG and
Emissions scores, firm size, and bank-debt share. The unconditional
ETI--bank-share correlation is 0.004
($N=13{,}182$; $p=0.626$). The TalkFreq variable in that table is the winsorized
presentation-plus-management-answer measure; it is distinct from the
presentation-only variable used in the ratio-free models.

Appendix Table~\ref{tab:reprisk_checks} compares the ratio-free configurations
with later RepRisk incidents as an outside check. The continuous
talk--emissions interaction has no statistically detectable association with
the logged greenwashing count
($p=0.371$), greenwashing incidence ($p=0.567$), or environmental incidents
($p=0.579$) in the firm-and-year models. The high-talk/rising-emissions category
has a 1.39-percentage-point coefficient for greenwashing incidence ($p=0.061$),
but the corresponding Holm value is 0.552. This is a directionally suggestive
raw association, not multiplicity-adjusted evidence that the construction
measures greenwashing.

\subsection{Call segments and subsequent environmental information}

Appendix Table~\ref{tab:segment_checks} reports the segment-level comparisons
with external outcomes.
The presentation- and answer-section measures enter jointly, so each
coefficient reports that segment's association conditional on the other segment measure.
The displayed columns include firm and year
effects, lagged firm controls, the prior emissions state, and the lagged value
of the outcome.

The presentation-section measure is positively associated with the
next-year count of externally recorded greenwashing incidents. A
one-standard-deviation increase in the transformed
prepared-talk measure is associated with a 0.0096 increase in
$\log(1+n)$ greenwashing incidents (standard error 0.0024, $p=0.002$;
$N=10{,}512$). The section-wide Holm value is 0.049. Answer talk has no statistically detectable
association with the same outcome ($\widehat\beta=-0.0011$, $p=0.665$).

The binary outcome points in the same direction but is weaker.  Prepared talk
is associated with a 0.75 percentage-point increase in the probability of at
least one next-year greenwashing incident (standard error 0.33 percentage
points, raw $p=0.044$). Its section-wide Holm value is 0.750. The count and
incidence results point in the same direction, but the latter does not remain
below conventional thresholds after the displayed adjustment.

The pattern is reversed for the broader environmental count. The answer-section measure is positively associated with next-year
$\log(1+n)$ environmental incidents after conditioning on the lagged count
($\widehat\beta=0.0107$, standard error 0.0030, raw $p=0.005$). Its
section-wide Holm value is 0.097.
The presentation-section measure is smaller and statistically indistinguishable
from zero ($\widehat\beta=0.0052$, $p=0.378$). In these models, the
presentation-section coefficient is statistically significant for the greenwashing-news count
but not the broader environmental count. Because news coverage and underlying
exposure can drive both, this remains an interpretation rather than an
identified mechanism.

Neither segment is statistically associated with subsequent emissions change
in the persistence model.
The prepared-talk coefficient is 0.0038 (standard error 0.0053, $p=0.498$) and
the answer-talk coefficient is 0.0027 (standard error 0.0030, $p=0.395$) in
8,388 observations. Table~\ref{tab:jfs_talk_emissions} separately reports the
relation between presentation talk and cumulative emissions changes over one,
two, and three years. The firm-effect coefficients are 0.0036, 0.0023, and
0.0002, with $p$-values of 0.382, 0.661, and 0.981. Presentation talk is also unrelated to later emissions
intensity (coefficient $-0.0016$, $p=0.757$). Thus,
we find no evidence
that more climate language is systematically followed by either emissions
abatement or increased emissions.

The segment results clarify what the fields capture; they do not establish a
deception measure.  Prepared language is positively associated with later externally
recorded greenwashing news, but not with emissions change; answer language is
associated with broader environmental news, but not greenwashing. The
information source and the definition of the outcome
matter. Prepared emphasis is associated with later negative-news
records and may accompany latent exposure not captured by the controls. It would
be an overreach to infer that the observed prepared passages contain false
claims or that talk causes the later incident.

\subsection{The ETI and debt composition}

Table~\ref{tab:jfs_baseline} reports the fixed-effect and timing comparisons;
Appendix Table~\ref{tab:jfs_exact_lag_components} gives the exact-lag results
for all three debt shares.
The first columns use the
industry-and-year design. With lagged log emissions but no other controls, the
ETI coefficient is $-0.2265$ (standard error 0.0843; $p=0.023$; $N=13{,}182$).
In the contemporaneous  specification, adding the six financial controls
reduces it to $-0.0914$ (standard error
0.0380; $p=0.037$; $N=10{,}691$). This full industry-effect estimate remains
economically modest: one standard deviation of ETI corresponds to roughly 0.4
percentage points of bank-debt share,
compared with a mean share near 34 percent.

The firm-and-year model changes the result. On its available sample, the
coefficient is $-0.0178$ (standard error 0.0517; $p=0.738$; $N=10{,}483$).
We also estimate both fixed-effect structures on exactly these 10,483
observations. The industry-and-year coefficient is then $-0.0961$ (standard
error 0.0332; $p=0.016$), while the firm-and-year coefficient is unchanged at
$-0.0178$. The reduction of about 81.5 percent therefore does not arise from
sample composition or the observations dropped as firm-effect singletons. It
arises when the comparison shifts from differences across firms within
industries to changes within the same firm.

Appendix Table~\ref{tab:jfs_within_between} makes the same distinction through
a within--between decomposition on the 10,483-observation sample. The
coefficient on the within-firm deviation is $-0.0256$ ($p=0.580$), while the
coefficient on the firm mean is $-0.3068$ ($p=0.222$). Both are imprecise, but
the contrast is consistent with the fixed-effect comparison: the data contain
more of the negative relation in persistent differences across firms than in
year-to-year changes within a firm.

The progression of controls is also informative. On the full industry-effect
sample, an ETI-only model gives a coefficient of $-0.0198$ ($p=0.746$). Adding
lagged emissions makes the estimate $-0.2328$ ($p=0.004$); adding the six
financial controls then reduces its magnitude to $-0.0914$. In the firm-effect
sample, the corresponding coefficients are $-0.0227$ ($p=0.606$), $-0.0512$
($p=0.320$), and $-0.0178$ ($p=0.738$). Thus the negative industry-effect
estimate is not an unconditional relation between ETI and bank share. It
appears after conditioning on emissions and is then materially reduced by the
financial controls. This pattern does not make the full model improper, but it
shows why coefficient stability cannot be summarized by the final column alone.

Table~\ref{tab:jfs_debt_components} shows that the public-debt share offers no
matching firm-effect result. In the
industry-and-year model, ETI has a positive bond-share coefficient of 0.1234,
but it is statistically imprecise ($p=0.125$). With firm and year effects the
coefficient is 0.0028 ($p=0.966$). We report the residual other-debt share here;
total-debt checks appear in Appendix Table~\ref{tab:jfs_total_debt}.
Because the three shares sum to
one, a change in one category must be considered with the others. None of these
regressions identifies a public bond issue or demonstrates improved access to
the public market.

Timing provides a separate check. The variable $ETI_{it}$ uses emissions growth
ending in $t$ and talk from $t-1$. Relating it to bank share in $t$ is therefore
partly contemporaneous. When the completed index is lagged one more calendar
year, the industry-and-year coefficient is $-0.0163$, corresponding to a one-
standard-deviation effect of $-0.0007$, on a common 8,334-observation sample.
Its unrepaired covariance has a positive diagonal and yields $p=0.374$, but the
two-way covariance matrix is not positive semidefinite. The standard eigenvalue
repair is sensitive to regressor scaling: estimating the raw-coordinate model
and then rescaling its coefficient algebraically gives $p=0.536$, while
standardized-coordinate implementations give values around 0.621--0.627. The
firm-and-year coefficient is 0.0371,
corresponding to a one-standard-deviation effect of 0.0016 ($p=0.606$).
Appendix Table~\ref{tab:jfs_exact_lag_higher_fe} reports exact-lag models that
pair firm effects with two-digit SIC-by-year or state-by-year effects. The predictive
specifications supply no evidence that ETI precedes a change in debt
composition; this does not establish exogeneity.

Appendix Table~\ref{tab:jfs_oster_itcv} reports the Oster and Frank
calculations, while Table~\ref{tab:jfs_inference} reports the covariance checks.
The Oster result depends sharply on the assumed maximum
$R^2$. In the firm-and-year model, holding lagged emissions in both the short
and full specifications gives an exact signed $\delta$ of 1.626 when
$R_{\max}=1.3R^2_{full}$, but 0.0043 when $R_{\max}=1$, using 10,483
observations. The corresponding industry-and-year values are 1.218 and 0.0616
on that model's 10,691-observation sample. This variation is unsurprising in
the firm-effect model because its within $R^2$
only rises from 0.00033 in the short model to 0.00867 in the full model, and the
1.3 rule sets $R_{\max}$ to just 0.01126. An ETI-only short model also produces
signed values that change materially with the rule, including a negative value
in the industry design because adding controls moves the coefficient away from
zero. We report the full grid rather than selecting a single threshold.

Applying Frank's published-results formula to the two-way-clustered
$t$-statistic, but using conventional regression degrees of freedom, gives a
plug-in impact threshold of approximately 0.0044. This corresponds to equal
conditional partial correlations of about 0.066 between an omitted variable
and ETI and between that variable and bank share. A cluster-calibrated analogue
using the year-cluster degrees of freedom is 0.0699, with equal partial
correlations of about 0.264. Neither is an exact extension of the method to a
two-way covariance. Under the homoskedastic OLS standard error for which the
calculation was derived, the coefficient already has $p=0.169$, so the orthodox
threshold needed to make it statistically insignificant is zero. We treat the
plug-in numbers as descriptive sensitivity statistics rather than corrections
for omitted variables.

The covariance estimates explain why the $p$-values differ. For the industry-and-year
coefficient, the homoskedastic, firm-clustered, year-clustered and two-way
clustered standard errors are 0.0664, 0.0697, 0.0551 and 0.0380. Their two-sided
$p$-values are 0.169, 0.190, 0.128 and 0.037. The two-way estimate is smaller
because the firm--year intersection term subtracts most of the sum of the two
one-way variance components. An exhaustive null-imposed year-sign bootstrap
gives $p=0.250$ when studentized with the year-cluster covariance and $p=0.122$
when studentized with the two-way covariance. In leave-one-year-out models the
industry coefficient is always negative but is below 5 percent in six of eleven
cases; no firm-effect estimate is below 10 percent. These checks do not show
that the industry-and-year coefficient is exactly zero. They show that it does
not provide a secure basis for a causal monitoring account.

\subsection{Ratio-free climate information and debt composition}

Table~\ref{tab:ratiofree_debt_main} presents the principal financing results.  The
continuous model includes standardized presentation talk, standardized
emissions change, and their product simultaneously.  With firm and year effects,
the bank-share coefficients are 0.0059 for emissions change (standard error
0.0031, $p=0.092$), $-0.0028$ for presentation talk (standard error 0.0031,
$p=0.386$), and $-0.0002$ for their interaction (standard error 0.0011,
$p=0.865$).  The three terms are jointly insignificant ($p=0.307$), and the
Holm values across the three debt outcomes are at least 0.828.  The estimation
sample contains 8,219 observations for 1,548 firms over ten outcome years.

The quadrant specification in Appendix Table~\ref{tab:quadrant_debt} reaches
the same conclusion without imposing a linear interaction.  Relative to low
talk/non-rising emissions, the
high-talk/rising-emissions coefficient for bank share is 0.0101 (standard error
0.0072, $p=0.193$).  The high-talk/non-rising and low-talk/rising coefficients
are also positive and individually insignificant, and the quadrant indicators
are jointly insignificant ($p=0.407$). The high-talk/rising-emissions estimate
therefore has the opposite sign from monitoring avoidance, but it is too
imprecise to support a positive financing claim.

Bond and other-debt shares do not supply a coherent offset.  There are isolated
raw estimates, including a negative other-debt coefficient for the low-talk/
rising-emissions category, but they are insignificant after the relevant family-wise
adjustment.  In the industry comparison, some quadrant coefficients for bond
share are positive, while the same-sample firm-effect estimates are smaller and
insignificant.  Given that the three shares sum to one, selective emphasis on a
single raw coefficient would be misleading. The component and
quadrant models do not show that talk--emissions configurations
predict a firm's subsequent debt mix.

Appendix Table~\ref{tab:jfs_ratio_tail_diagnostics} reports reconstructed
presentation-only and change-on-change ratios before and after fixed tail
treatment. Their
estimates change in magnitude and sometimes in direction. Together with the
correctly timed ETI results in Table~\ref{tab:jfs_baseline}, this shows that the
negative coefficient is confined to its contemporaneous industry-
effect specification rather than appearing across reasonable ways of joining
talk and emissions.

\section{Exploratory heterogeneity}\label{sec:heterogeneity}

We examine whether the ETI association varies with product-market competition,
financial health, information opacity, and the Paris Agreement period. An
interaction in an observational debt-share regression
does not identify monitoring avoidance. It may reflect borrower demand, lender
supply, public-market access or another characteristic correlated with the
moderator. We therefore describe these as exploratory comparisons, show both
fixed-effect structures, and consider the number of related tests.

 We do not report two further moderator splits. A state-political
classification  would raise the same state-level confounding
concern as the  instrument discussed above, and a bank-debt-level split
 conditions on a close relative of the outcome.

\subsection{Product-market competition}

A possible competition argument is that firms in more competitive markets
may face stronger short-run pressure and therefore place greater weight on
climate communication relative to operational change. A monitoring-avoidance
account would then predict greater bank scrutiny and less borrower demand for
bank debt. The first step is possible but not measured directly; the second
cannot be inferred from an annual debt share. Product-market competition is
itself related to debt choice and managerial incentives
\citep{boubaker2018,karuna2007}; these links need not operate through climate
discussion.

Appendix Table~\ref{tab:jfs_competition} reports the competition tests.
Low product substitutability, high industry concentration and a low number of
competitors are defined relative to the sample median within the available
variables. Each moderator, ETI and their interaction enter with the baseline
controls. Under industry and year effects, the interaction coefficients are
positive: 0.3035 for low substitutability, 0.3365 for high concentration and
0.4155 for few competitors. Their $p$-values are 0.067, 0.007 and 0.004.
These estimates imply that the negative ETI coefficient is weaker in less
competitive settings.

With firm and year effects, the same interactions fall to 0.1204, 0.0738 and
0.1702, with $p$-values of 0.348, 0.524 and 0.124. None is below the 5-percent
threshold, and none provides strong evidence after adjustment across the
competition family. In the exact-lag models reported in Appendix
Table~\ref{tab:jfs_clean_exact_lag_interactions}, all six competition
interactions have raw $p$-values above 0.54. The comparison suggests that the
contemporaneous pattern largely reflects persistent differences across firms
or industries.

\subsection{Financial health}

Financial condition can affect debt choice through several routes. A safer
borrower may obtain public debt on better terms, but it may also retain bank
relationships at lower cost. A constrained firm may value renegotiation and
relationship capital, while a distressed firm may face restricted credit
supply. Financial health can also affect environmental investment and the
volatility of emissions. These competing channels make the interaction
descriptive rather than a direct test of monitoring intensity.

Appendix Table~\ref{tab:jfs_financial_health} considers dividend payment, an
above-median Altman $Z$-score and above-median market capitalization. The
industry-effect interactions are positive, but none is individually
statistically significant. With firm effects, the
ETI--dividend interaction is 0.1766 ($p=0.255$), the ETI--high-market-value
interaction is $-0.0228$ ($p=0.795$), and the ETI--high-$Z$ interaction is
0.2091 ($p=0.065$). None crosses the 5-percent threshold even before accounting
for the three related moderators, and the estimates sit beside statistically
insignificant main ETI coefficients. We do not interpret the high-$Z$ estimate
as evidence that financial health weakens bank monitoring. In the exact-lag
models, the firm-effect dividend and market-capitalization interactions have
raw $p$-values of 0.060 and 0.043, respectively, but both have within-family
Holm values of 0.128 and their signs differ. This is not a common financial-
health pattern.

\subsection{Information opacity}

Opacity is theoretically important but does not have a single implication for
bank debt. Banks have an advantage over arm's-length investors in producing
private information \citep{fama1985,krishnaswami1999}, which may make bank debt
more valuable to opaque firms. Poorer accounting quality is associated with
more stringent private-debt terms \citep{bharath2008}, while an informational
advantage can strengthen lender bargaining power \citep{rajan1992}. Changes in
the information environment can also alter firms' financing choices
\citep{li2019}.
We consider four proxies: analyst forecast dispersion, low analyst coverage, a
high  bid--ask proxy, and high intangible intensity. These measures
reflect different aspects of the information environment and none is treated
as universally preferable.

Appendix Table~\ref{tab:jfs_opacity} reports all four interactions. Under firm and
year effects, the interaction $p$-values are 0.275 for forecast dispersion,
0.030 for low coverage, 0.272 for the  bid--ask proxy and 0.897 for
intangibles. The low-coverage estimate is positive, but its Holm value across
the four opacity tests is 0.120. The other three estimates are individually
imprecise. We therefore do not find a consistent opacity pattern. The result
also illustrates why the choice of a single proxy after inspection would be
misleading: the only raw coefficient below 5 percent is not the  bid--ask proxy. Rebuilding the moderator dummies so missing source observations
remain missing and imposing exact timing changes the pattern. The high-intangibles
interaction is positive with firm effects and has a raw $p$-value of 0.0014 and
a within-opacity Holm value of 0.0055. Its two-way covariance requires an
eigenvalue correction that is sensitive to scaling; an exact year-sign check
has $p=0.043$ before and 0.391 after adjustment across all ten exact-lag moderator
tests. We therefore report the estimate but do not generalize it across the
four opacity proxies.

\subsection{Time pattern and the Paris Agreement}

We divide the sample into years before 2016 and years from 2016 onwards. The split
also coincides with changes in green-bond markets, environmental reporting,
investor demand, regulation and the composition of firms covered by the linked
databases. It is therefore not a Paris Agreement research design.

Appendix Table~\ref{tab:jfs_paris} retains the split. With industry and year effects,
the pre-2016 ETI coefficient is $-0.0079$ ($p=0.980$) and the post-2016
coefficient is $-0.0949$ ($p=0.032$). With firm and year effects they are
$-0.0599$ ($p=0.753$) and $-0.0168$ ($p=0.750$). The relevant question is not
whether one coefficient crosses a threshold while another does not, but whether
the two differ. The formal pre/post equality tests are statistically
insignificant under both structures ($p=0.783$ for industry effects and
$p=0.812$ for firm effects). The annual coefficients in
Figure~\ref{fig:year_specific_combined} also vary around zero and their
95-percent intervals overlap zero. These results do not support a discrete
change after 2015. Exact-lag versions lead to the same conclusion whether the
post period is dated by the debt-outcome year or the index source year; every
pre/post equality-test $p$-value exceeds 0.50.

The heterogeneity results do not establish a monitoring
mechanism. Several industry-effect patterns are present, but they do not form a
coherent within-firm pattern once firm effects, common samples and the number of
related tests are considered. This does not rule out variation in how firms and
banks respond to climate information. It means that these moderators cannot
identify that variation in the present panel.

\section{Additional analyses}\label{sec:additional}

\subsection{Comparison across information sources}

Table~\ref{tab:external_horse_race} asks whether another climate field adds
information about debt composition once the interaction and the other measures
are included. Ten variables enter simultaneously: presentation
talk, emissions change, their interaction, environmental incidents, a
greenwashing indicator, Sautner opportunity, regulatory, and physical
exposures, and LSEG ESG and Emissions scores.  The common sample contains 8,061
observations for 1,529 firms over ten years.

For bank-debt share, the talk coefficient is $-0.0024$ ($p=0.478$), the
emissions-change coefficient is 0.0049 ($p=0.156$), and their interaction is
$-0.0002$ ($p=0.829$).  Environmental incidents are also insignificant in this
restricted common sample ($-0.0023$, $p=0.281$), as are the greenwashing
indicator and both LSEG scores.  The Sautner physical-exposure coefficient is
0.0100 with a raw $p$-value of 0.039.  It is insignificant after adjustment across
the displayed information coefficients; all Holm values in the joint comparison are
one. The bank-share covariance required  a positive-semidefinite repair, so the isolated raw coefficient warrants additional caution. No
information measure remains significant in the analogous bond- and other-debt
families.

In this restricted common sample, none of the measures has a statistically
significant additional association with debt composition after conditioning on
the others. This
result does not imply that every
measure is uninformative in other settings, and it does not show that the
talk--emissions index measures the same object.

The segment-level results provide a useful contrast. Prepared talk has a
forward association with an external greenwashing count, yet it does not
explain the debt mix. A measure can be associated with an external outcome without lenders or
borrowers responding in a way detectable in annual debt shares.  Conversely,
the Sautner topical exposures and LSEG scores need not reproduce findings from
other samples or outcomes.  Exposure, rating coverage, score levels, and
negative news answer different questions.

\subsection{External controversies and bank-debt share}

Appendix Table~\ref{tab:jfs_reprisk} begins with externally recorded events
rather than managerial language.  In the model that includes standardized
environmental incidents and the greenwashing indicator, a one-standard-deviation
increase in the environmental count at $t-1$ is associated with a
0.61-percentage-point decrease in bank share at $t$ (coefficient $-0.0061$, standard error
0.0023, raw $p=0.023$; $N=10{,}373$). The greenwashing indicator has no
statistically detectable association with bank share (coefficient $-0.0016$,
$p=0.835$). In a separate model, a one-standard-deviation increase in total ESG
incidents is associated with a 1.13-percentage-point decrease in bank share
(coefficient $-0.0113$, standard error
0.0047, raw $p=0.039$).

The magnitude is modest relative to an average bank share near 34 percent, and
specificity checks narrow the interpretation. When environmental incidents are
conditioned on total incident activity and the greenwashing indicator, the
displayed environmental coefficient falls to $-0.0016$ ($p=0.550$). The broad
total-incident measure is associated with bank share when estimated separately,
whereas greenwashing-specific measures remain statistically insignificant.

The environmental estimate is suggestive before adjustment but not after
accounting for the testing family. The exactly enumerated one-way
year-cluster wild-bootstrap gives $p=0.024$. Yet the Holm value across the six
selected wild-bootstrap checks is 0.141, and the one-way bootstrap omits
within-firm dependence.  In the broad primary testing family, the adjustment is
larger still.

Appendix Table~\ref{tab:reprisk_dynamics} reports lead--lag estimates.
In a common 6,526-observation distributed model with firm and year effects, the
lagged environmental coefficient is $-0.0042$ ($p=0.227$), the current-event
coefficient is 0.0013 ($p=0.769$), and the future-event coefficient is
$-0.0062$ ($p=0.330$).  None is individually significant.  A descriptive
first-event model similarly fails to support a causal interpretation: although
individual within-firm pre-event coefficients are imprecise, their pre-event
joint test rejects at the 5 percent level.  Incident timing is therefore not a
credible quasi-experiment in this panel.

We conclude that broad controversy activity may contain financing information
that the internal talk--emissions measures do not.  The data do not establish
that the association is environmental rather than a manifestation of general
reputation or governance risk.  They also cannot distinguish firms choosing
less bank debt from banks supplying less credit.  The absence of a
greenwashing-specific coefficient is particularly important: the prepared-talk
result for subsequent news does not translate into a
greenwashing-specific bank-share result.

\subsection{Alternative measures and specification checks}

Appendix Table~\ref{tab:jfs_alternative_measures} reports three further
constructions. The first replaces presentation-plus-management-answer talk in
the ETI denominator with the presentation-only
frequency. This addresses the concern that an answer may be prompted by an
analyst, but it does not resolve the small-denominator problem or show that a
prepared passage contains a commitment. Its bank-share coefficient is just
above the 5-percent threshold in the industry-effect model ($p=0.053$) and statistically
insignificant with firm effects ($p=0.731$).

The second divides the change in log emissions by the change in presentation
talk. This places changes in both numerator and denominator, but there is no
economic reason to expect proportional changes in these two objects, and the
ratio is undefined when talk does not change. Its  raw distribution ranges from
approximately $-40.24$ million to $2.62$ million in the units used in the data. The industry-effect coefficient is positive ($p=0.013$), opposite to
the ETI sign, and becomes statistically insignificant with firm
effects ($p=0.818$). Winsorization changes its scale and can alter its sign.
Using changes in both numerator and denominator does not by itself make the
ratio economically preferable, so we treat it as a sensitivity check.

The third is an industry-year adjusted, standardized sum of presentation talk
and emissions intensity. It avoids division and is defined when talk is zero,
but it imposes an equal-weight aggregation whose economic meaning differs from
ETI. Its coefficient is statistically insignificant with both
industry effects ($p=0.539$) and firm effects ($p=0.485$). The component and
quadrant results reported earlier are easier to interpret than this composite
and therefore receive more weight.

We also compare the component measures with LSEG ESG and Emissions scores and
with the Sautner opportunity, regulatory, and physical exposure measures. The
ten-variable comparison in
Table~\ref{tab:external_horse_race} leaves every coefficient statistically
insignificant after the declared family-wise adjustment. These findings do not
show that the external measures are uninformative in general. They show that
ETI is not a simple substitute for them and that none has a statistically
significant additional association with annual debt shares after conditioning
on the other measures in the common sample.

As further score and common-shock checks, Appendix
Table~\ref{tab:jfs_archival_table13} adds current or lagged LSEG ESG and
Emissions scores and absorbs two-digit SIC-by-year effects. The industry-design
ETI coefficients remain negative in these specifications, whereas the
corresponding firm-design coefficients are small and statistically imprecise.
Industry-by-year effects do not replace firm effects; they absorb a different
set of common shocks. These columns do not change the conclusion that the
within-firm and predictive ETI estimates are statistically indistinguishable
from zero.

\subsection{DealScan evidence from unique facilities}

Table~\ref{tab:jfs_dealscan_spread} evaluates the ETI in the
one-record-per-facility file  over complete years 2010--2019,  holding the sample
fixed across fixed-effect designs  and clustering by firm and year. In the
contemporaneous comparison, the one-standard-deviation ETI coefficient for the
log all-in-drawn spread is 0.0473 ($p=0.115$) with industry and year effects
and 0.0039 ($p=0.883$) with firm and year effects. With the  index lagged by one
calendar year, the coefficients are 0.0054 ($p=0.749$) and $-0.0304$
($p=0.169$), respectively. All four estimates are statistically indistinguishable from zero, and
their magnitudes vary with timing and fixed effects.

Appendix Table~\ref{tab:jfs_dealscan_extensive} examines appearance and volume
in the linked extract. Industry-effect coefficients are positive for any
matched facility, facility count, and reported amount, with raw $p$-values of
0.023, 0.044, and 0.019. The corresponding firm-effect coefficients are near
zero, with $p$-values of 0.942, 0.785, and 0.928. These outcomes measure what is
linked in DealScan; a non-match does not mean that a firm obtained no bank
credit elsewhere.

The accepted-facility terms do not yield a matching pattern. Appendix
Table~\ref{tab:jfs_dealscan_terms} reports firm-effect coefficients of 0.018
($p=0.334$) for the number of financial covenants, $-0.014$ ($p=0.138$) for
security, and 0.007 ($p=0.544$) for log maturity. These contemporaneous
specifications use firm-clustered standard errors and the available contract
sample. The ratio-free facility tests in Appendix
Tables~\ref{tab:dealscan_facility_main} and
\ref{tab:dealscan_contract_robustness} are also mixed: the continuous
interaction is statistically insignificant for spread, security, and
covenants, while isolated maturity and revolver coefficients do not remain
below conventional thresholds after adjustment across the displayed facility
outcomes. A high-talk/rising-emissions category likewise gives no common
firm-effect pattern across incidence, pricing, maturity, security, and
covenants.

Lagged RepRisk environmental incidents and the greenwashing indicator also fail
to yield a coherent contract pattern. In the firm-effect models, neither
measure is statistically associated with spread, maturity, security, covenants,
revolver status, or facility incidence after the relevant outcome-family
adjustment (Appendix Tables~\ref{tab:dealscan_facility_main} and
\ref{tab:dealscan_firmyear}). Supplementary controls for loan type, deal
purpose, and currency, and a U.S.-dollar subsample, do not reveal a common
adverse-pricing relation (Appendix
Table~\ref{tab:dealscan_contract_robustness}).

These findings should not be read as evidence that banks ignore climate
information.  The extract has neither rejected applications nor lender
identities, lender environmental characteristics, green-loan flags, or complete
bilateral lending. It also conditions facility-term equations on selection
into an accepted syndicated deal. The available loan outcomes do not support a
consistent screening, monitoring-avoidance, or favorable-access interpretation
of the annual talk--emissions measures.

\subsection{Total debt, annual stability, and further inference checks}

Appendix Table~\ref{tab:jfs_total_debt} tests whether the share results
mask changes in the total-debt denominator. The talk--emissions interaction has
no statistically detectable association with log total debt (coefficient
0.0023, $p=0.651$) or rebuilt total debt-to-assets (coefficient $-0.0001$,
$p=0.916$). The high-talk/rising category also has no statistically detectable
association with either outcome. Environmental incidents have a raw
coefficient of 0.0047 on debt-to-assets ($p=0.054$), while their
log-total-debt coefficient is 0.0038 ($p=0.696$). The
greenwashing indicator is insignificant for both. These regressions do not
identify component amounts, but they provide no evidence that a consistent
total-borrowing response underlies the statistically insignificant debt-share estimates.

Figure~\ref{fig:year_specific_combined} examines year-specific stability. Panel A
plots annual industry-effect coefficients for contemporaneous ETI
with two-digit industry effects; firm effects are undefined within a one-year
cross-section. Panel B plots descriptive year-specific slopes for the
ratio-free interaction in the pooled firm-and-year specification with firm-clustered
confidence intervals. The estimates vary around zero, and every 95 percent confidence
interval includes zero. The ETI estimate is most negative in 2014
and has its smallest raw $p$-value in 2017 ($p=0.065$), but no annual 95 percent
confidence interval excludes zero. The ratio-free interaction likewise shows no stable
pre/post-2016 change.  The figure is not a Paris Agreement event study and does
not support a structural-break claim.

Table~\ref{tab:jfs_inference} and Appendix
Table~\ref{tab:jfs_wild_protocol} bring together the sensitivity results. The exact
year-cluster wild-bootstrap $p$-values are 0.852 for the
interaction and bank share, 0.117 for the high-talk/rising bank-share contrast,
0.024 for environmental incidents and bank share, 0.045 for high-talk/rising
emissions and future greenwashing incidence, 0.242 for the interaction and
facility spread, and 0.023 for high-talk/rising emissions and facility count.
None is significant after Holm adjustment across the six checks; the smallest
adjusted value is 0.141. The bootstrap results support the same distinction:
the central talk--emissions financing terms remain statistically
insignificant, while a
few external or secondary outcomes remain suggestive but cannot support broad
claims.

\section{Discussion}\label{sec:discussion}

\subsection{What the comparison reveals}

The results show that climate information is source-dependent. The
presentation- and answer-section measures are positively related to different
future RepRisk categories. The two segment measures are not similarly related
to later emissions changes, and none of the internal talk--emissions configurations is
consistently related to debt composition. Broad external controversy counts exhibit a modest
bank-share relation that is not greenwashing-specific. The differences indicate
that words, operational outcomes, vendor scores, topical exposures, and
negative news capture different information sets.

This distinction helps reconcile our evidence with recent work. Studies of climate exposure
report greater bank reliance and interpret it as consistent
with the value of flexibility and monitoring \citep{alayyoub2026}; studies of negative
ESG news can find a shift toward public bonds \citep{newton2024}.  A climate
word-frequency interaction need not reproduce either result.  Similarly,
evidence that adverse climate news affects loan terms \citep{hrazdil2024} does
not imply that prepared talk or an emissions ratio will do so.  The empirical
content of a variable follows from its construction and source, not the breadth
of the label ``climate risk.''

One exploratory result is the positive association between the presentation-
section measure and the later count of greenwashing news, but it
must be interpreted carefully. The measure precedes
the count of greenwashing news even after prior incidents and the answer-section
measure are included. This may reflect information common to the presentation
field and later negative-news coverage. It is also consistent with omitted exposure,
visibility, stakeholder attention, or contemporaneous initiatives that raise
both prepared discussion and later news coverage.  Without the underlying
sentences and commitments, the result cannot distinguish a misleading claim
from legitimate discussion of a difficult transition.

The answer-section result calls for the same caution. If the segment assignment is
accurate, the field partly reflects an interactive exchange rather than a purely
firm-initiated script. Its relation with broad environmental incidents may arise
because already-exposed firms receive more attention. It would be inappropriate
to describe the answer-section measure as independently verified
voluntary disclosure.

\subsection{Implications for debt-choice interpretation}

Introducing firm effects changes both the ETI estimate and its debt-choice
interpretation. The industry-effect coefficient compares firms
that differ persistently
in technology, disclosure, market access, and financing policy.  Firm effects
ask whether a given firm's later debt share changes when its own information
measure changes.  The latter estimate is near zero for the ETI and
for the ratio-free alternatives. It is therefore not defensible to infer
that managers dynamically avoid bank scrutiny from the industry-and-year
coefficient.

The sign ambiguity is not merely econometric.  Monitoring can be costly to
borrowers, but it can also be valuable when risk is complex and renegotiation is
important.  Bank screening and borrower avoidance produce the same negative
share sign, while monitoring value can produce the opposite sign.  Our loan
tests were designed to seek additional observable implications, yet their
isolated maturity, revolver, and facility-count estimates do not form a consistent pattern. A
realized share plus mixed accepted-loan terms cannot identify which side of the
market moved.

External controversies may matter differently because they are public negative
signals rather than combinations of a firm's own words and a vendor emissions
field.  The separate environmental and total-incident estimates resemble the
direction documented for broader reputation risk \citep{newton2024}.  However,
their overlap and lack of greenwashing specificity, together with the weak
lead--lag results for the environmental estimate, prevent a claim that banks respond
to a particular environmental-controversy shock.  General
reputation risk, business deterioration, public-market activity, or changing
financing demand can all produce the association.

\subsection{Relevance for financial stability}

The financial-stability implication is about information quality and model
risk, not a demonstrated systemic effect.  Credit institutions and supervisors
often face pressure to summarize heterogeneous climate information in tractable
scores.  Ratios are attractive because they appear to compare words with
outcomes. Here, small denominators create enormous values, tail treatment
changes results, and the negative industry-and-year association loses statistical
significance with an exact one-calendar-year lag and within-firm estimation.  A mechanically constructed
score can therefore look precise while remaining economically ambiguous.

Our evidence does not establish that the observed associations affect bank
losses, fire-sale risk, credit supply over a cycle, or aggregate stability. It
illustrates a measurement-risk concern for institution-level risk assessment:
a volatile measure could be used to sort borrowers or enter stress models even when its
within-firm financing content is negligible. Assessing such measures requires documented units and timing, attention to
the differences between sources, denominator checks, and an assessment of
out-of-sample or forward predictive content before interpreting them as risk
measures.

\section{Limitations}\label{sec:limitations}

Six limitations define the scope of the conclusions. First, the available data
contain annual precomputed transcript measures rather than raw calls
and upstream extraction code.  We can distinguish presentation and
answer segments, preserve zeros, and state our subsequent transformations.  We
cannot verify every speaker classification, bigram denominator, quarterly
weight, or omitted call, nor can we classify sentence valence.  The segment
associations should therefore be interpreted at the level of the available
fields.

Second, the LSEG/Refinitiv emissions panel does not provide complete
reported-versus-estimated flags, original publication dates, or historical
vintages.  Scope-specific coverage is sparse relative to the total field, and
exactly unchanged annual totals are common.  Calendar lags prevent use of future
rows, but they cannot ensure that every vendor value was available in
the same form at the relevant historical date.

Third, RepRisk incidents are event-based negative-news measures.  Media
visibility, geography, firm size, and provider coverage can affect recording.
The greenwashing label is closer to the construct of misleading environmental
communication than a talk--emissions ratio, but it is still neither a court
finding nor a complete census.  Predicting a RepRisk count can reflect future
scrutiny as well as underlying conduct.

Fourth, debt shares do not retain the associated bank, bond, and other-debt
amounts.  We cannot identify green-bond issuance, sustainability-linked loans,
ratings, or issuance opportunities. Although the total-debt checks help, a
share movement cannot be assigned to a numerator or denominator component.
The sample comprises public firms covered by the linked
databases: the location field is incomplete and includes Canadian
provinces, so the data do not support an exclusively U.S.-domiciled claim.

Fifth, the DealScan data observe accepted syndicated facilities, not
applications, withdrawn deals, rejections, or all bilateral credit.  They lack
lender identifiers and lender climate characteristics in the DealScan extract.
The file is incomplete after mid-July 2020 and absent in 2021, and several
contract fields sharply reduce the usable sample. Unknown collateral status is
left missing in the principal loan analyses. These features prevent us from separating demand from supply.

Sixth, the panel has only eight to eleven outcome-year clusters in key models.
Two-way clustering, exact year-cluster wild-bootstrap enumeration,
leave-one-year-out estimates, and family-wise adjustments expose this weakness
but do not create additional independent time variation.  Several raw results
are sensitive to the testing family, and the one-way bootstrap cannot
simultaneously model firm dependence.  All estimates are conditional
associations; none should be interpreted as a causal effect of talk, emissions,
or incidents on financing.

These limitations restrict what can be learned from the available data. Within
those limits, the paper reports statistically insignificant within-firm
estimates, checks on the denominators, differences between the two call
segments, and differences between internal and external measures.

\section{Conclusion}\label{sec:conclusion}

This paper asks which forms of climate information are associated with
corporate debt composition across the reported specifications. The answer depends sharply on what is
measured. In exploratory tests, the presentation-section measure has a small
positive association with the next-year count of externally recorded greenwashing news,
while the answer-section measure has different content for broader
environmental incidents. Neither segment measure is statistically associated
with later emissions growth. The segment fields have limited predictive
associations with later RepRisk incidents, but word frequency is not a promise, a costly signal,
or proof of misleading conduct.

Outside the contemporaneous industry-and-year ETI benchmark, the principal
financing estimates are statistically insignificant.
Presentation talk,
emissions change, their interaction, and complete talk--emissions quadrants do
not consistently predict subsequent bank, bond, or other-debt shares within firms.
The negative ETI relation appears only in the contemporaneous industry-effect
specification; it is statistically insignificant with firm effects and is not
recovered when the completed index is correctly lagged.  Its
one-standard-deviation magnitude in the former model is about 0.4 percentage
points.  Ratio tails and trimming
sensitivity explain why ratio-free constructions should carry the main weight.

External controversies contain somewhat different information.  Lagged
environmental and total RepRisk counts have modest negative bank-share
associations in separate raw tests. The result is not greenwashing-specific;
the environmental association loses magnitude and statistical significance
when total incidents enter, and
wider-family multiplicity adjustment weakens both associations. The lead--lag
estimates further weaken the environmental result. DealScan outcomes provide no
consistent accompanying pattern in facility incidence, pricing, maturity,
security, covenants, or revolver status.  The data therefore do not establish
that borrowers avoid monitoring or that banks ration firms on the basis of the
measures studied here.

For research and financial-stability practice, the evidence calls for careful
measurement. Earnings-call language, operating emissions, topical
exposure, vendor scores, and external negative news should not be collapsed
under a common credibility label without construct-specific evidence.
Explicit timing, within-firm comparisons, denominator checks, and
outcome-family inference materially change the conclusion.  Better evidence on
sentence-level commitments, historical-vintage emissions, issue-level debt,
loan applications, and lender characteristics may reveal financing responses
that the present data cannot identify. With the information available here,
estimated associations with debt composition depend on the source and
construction of climate information, and ETI has
no consistent incremental association with debt composition.

\clearpage
\section*{Tables}


\providecommand{\jfsresult}[3]{\shortstack{#1\\{\scriptsize (#2)}\\[-0.35ex]{\scriptsize $p=#3$}}}
\providecommand{\jfsdash}{\multicolumn{1}{c}{---}}

\begin{table}[!htbp]
\centering
\caption{Sample construction and principal variables}
\label{tab:jfs_sample}
\label{tab:sample_construction}
\label{tab:descriptive_statistics}
\begin{threeparttable}
\scriptsize
\begin{adjustbox}{max width=\textwidth}
 \begin{tabular}{@{}lrrrrrr@{}}
\toprule
 & $N$ & Firms & Mean & SD & Median & Coverage / window \\
\midrule
\multicolumn{7}{@{}l}{\textit{Panel A: files and regression samples}}\\
Firm--year master panel & 16,258 & 2,707 & & & & 2010--2021 \\
Firm--years with bank, bond and other debt shares & 15,912 & 2,694 & & & & 2010--2021 \\
Full bank-share specification & 10,691 & 2,034 & & & & 2011--2021 \\
Same-sample firm/year specification & 10,483 & 1,826 & & & & 2011--2021 \\
Exact-lag ETI specification & 8,334 & 1,560 & & & & 2012--2021 \\
Unique-facility DealScan file & 15,958 & 2,631 & & & & 2010--2020 \\
Complete-year spread sample, contemporaneous index & 2,507 & 525 & & & & 2010--2019 \\
Complete-year spread sample, exact-lag index & 1,988 & 433 & & & & 2010--2019 \\
\addlinespace
\multicolumn{7}{@{}l}{\textit{Panel B: firm--year variables}}\\
Bank debt share & 15,912 & & 0.341 & 0.362 & 0.200 & 2010--2021 \\
Public bond debt share & 15,912 & & 0.498 & 0.390 & 0.553 & 2010--2021 \\
Other debt share & 15,912 & & 0.160 & 0.281 & 0.026 & 2010--2021 \\
Emissions-to-Talk Index & 13,410 & & 0.0030 & 0.0439 & 0.0000 & 2010--2021 \\
Presentation climate-term frequency $\times 10{,}000$ & 16,040 & & 10.474 & 27.144 & 2.607 & 2010--2021 \\
Log Scope 1+2+3 emissions & 16,258 & & 12.753 & 1.952 & 12.691 & 2010--2021 \\
Annual change in log emissions & 13,410 & & 0.024 & 0.247 & 0.000 & 2011--2021 \\
Log assets & 16,070 & & 8.448 & 1.855 & 8.399 & 2010--2021 \\
Total debt/assets & 16,022 & & 0.298 & 0.234 & 0.271 & 2010--2021 \\
PP\&E/assets & 15,353 & & 0.262 & 0.257 & 0.163 & 2010--2021 \\
Net income/assets & 16,069 & & 0.016 & 0.186 & 0.034 & 2010--2021 \\
Tobin's Q & 15,859 & & 2.063 & 1.533 & 1.538 & 2010--2021 \\
Altman Z-score, winsorized & 13,254 & & 1.308 & 1.853 & 1.442 & 2010--2021 \\
\bottomrule
\end{tabular}
\end{adjustbox}
\begin{tablenotes}[flushleft]\footnotesize
\item \textit{Notes:} The  master panel contains 16,258 firm--years. Debt shares are proportions and sum to one, subject to rounding. ``ETI'' denotes the  directional ratio of the contemporaneous emissions change to prior climate-talk frequency. The  unique-facility DealScan file contains 15,958  facilities. Exact calendar lags are set to missing across panel gaps. For the two complete-year spread rows, 2010--2019 is the analysis window; the fitted samples contain nine and eight outcome years, respectively.
\end{tablenotes}
\end{threeparttable}
\end{table}

\begin{table}[!htbp]
\centering
\caption{Climate talk and subsequent emissions}
\label{tab:jfs_talk_emissions}
\label{tab:talk_future_emissions}
\begin{threeparttable}
\scriptsize
\begin{adjustbox}{max width=\textwidth}
\begin{tabular}{@{}lcccccc@{}}
\toprule
 & \multicolumn{3}{c}{Industry + year effects} & \multicolumn{3}{c}{Firm + year effects} \\
\cmidrule(lr){2-4}\cmidrule(l){5-7}
Talk measure or horizon & Estimate & $N$ & Firms & Estimate & $N$ & Firms \\
\midrule
\multicolumn{7}{@{}l}{\textit{Panel A: log emissions at $t$; talk measured at $t-1$}}\\
Presentation and management-answer portions
 & \jfsresult{2.7525}{1.0068}{.021} & 10,682 & 2,114
 & \jfsresult{$-0.1160$}{0.2859}{.693} & 10,454 & 1,886 \\
Prepared-presentation portion
 & \jfsresult{37.8280}{13.6385}{.020} & 10,671 & 2,114
 & \jfsresult{1.2744}{4.2558}{.771} & 10,443 & 1,886 \\
\addlinespace
\multicolumn{7}{@{}l}{\textit{Panel B: cumulative change in log emissions from $t$ to $t+h$; presentation talk measured at $t$}}\\
$h=1$
 & \jfsresult{0.00235}{0.00247}{.364} & 10,673 & 2,020
 & \jfsresult{0.00355}{0.00389}{.382} & 10,477 & 1,824 \\
$h=2$
 & \jfsresult{0.00193}{0.00322}{.563} & 8,634 & 1,825
 & \jfsresult{0.00233}{0.00514}{.661} & 8,371 & 1,562 \\
$h=3$
 & \jfsresult{0.00052}{0.00438}{.908} & 6,803 & 1,565
 & \jfsresult{0.00015}{0.00600}{.981} & 6,563 & 1,325 \\
\addlinespace
\multicolumn{7}{@{}l}{\textit{Panel C: log emissions intensity at $t$; presentation talk measured at $t-1$}}\\
Log emissions intensity
 & \jfsresult{0.00368}{0.00231}{.142} & 10,424 & 1,809
 & \jfsresult{$-0.00156$}{0.00492}{.757} & 10,424 & 1,809 \\
\bottomrule
\end{tabular}
\end{adjustbox}
\begin{tablenotes}[flushleft]\footnotesize
\item \textit{Notes:} Each estimate is followed by its two-way firm/year-clustered standard error and unadjusted two-sided $p$-value. Panel A relates current log emissions to prior-year talk and uses rebuilt controls. TalkFreq for the combined presentation and management-answer portions is scaled by ten, while presentation-only talk is a raw fraction, so their coefficients should not be compared in magnitude. Panels B and C standardize presentation talk within two-digit-industry/year cells and control for the prior outcome, assets, debt/assets, PP\&E/assets, profitability, Tobin's Q and Altman Z-score. Panel B outcomes require an exact $h$-year calendar match. Panel A has 11 year clusters. Panel B has 11, 10 and 9 year clusters for $h=1,2,3$, respectively; Panel C has 11. Both firm/year covariances in Panel A and the industry/year covariance for Panel B at $h=1$ required  a positive-semidefinite repair; no other displayed covariance did. The firm-effect estimates do not indicate that greater presentation talk precedes lower emissions or emissions intensity over these horizons.
\end{tablenotes}
\end{threeparttable}
\end{table}

\begin{table}[!htbp]
\centering
 \caption{Emissions-to-Talk Index and bank debt share}
\label{tab:jfs_baseline}
\label{tab:eti_baseline}
\begin{threeparttable}
\scriptsize
\begin{adjustbox}{max width=\textwidth}
 \begin{tabular}{@{}lccccc@{}}
\toprule
 & \multicolumn{2}{c}{Baseline sample} & \multicolumn{2}{c}{Common sample} & Exact lag \\
\cmidrule(lr){2-3}\cmidrule(lr){4-5}\cmidrule(l){6-6}
 & Industry/year & Industry/year & Industry/year & Firm/year & Firm/year \\
 & parsimonious & full controls & full controls & full controls & full controls \\
\midrule
ETI
 & \jfsresult{$-0.2265$}{0.0843}{.023}
 & \jfsresult{$-0.0914$}{0.0380}{.037}
 & \jfsresult{$-0.0961$}{0.0332}{.016}
 & \jfsresult{$-0.0178$}{0.0517}{.738}
 & \jfsresult{0.00159}{0.00297}{.606} \\
Lagged log emissions included & Yes & Yes & Yes & Yes & Yes \\
Six lagged firm controls & No & Yes & Yes & Yes & Yes \\
Index timing & $t$ & $t$ & $t$ & $t$ & $t-1$ \\
Index scaling & raw & raw & raw & raw & one SD \\
Observations & 13,182 & 10,691 & 10,483 & 10,483 & 8,334 \\
Firms & 2,476 & 2,034 & 1,826 & 1,826 & 1,560 \\
Year clusters & 11 & 11 & 11 & 11 & 10 \\
$R^2$ & 0.171 & 0.234 & 0.239 & 0.784 & 0.796 \\
Within $R^2$ & 0.065 & 0.144 & 0.148 & 0.009 & 0.003 \\
\bottomrule
\end{tabular}
\end{adjustbox}
\begin{tablenotes}[flushleft]\footnotesize
\item \textit{Notes:} The dependent variable is bank debt divided by total debt. The  ETI for year $t$ already combines the emissions change ending in $t$ with talk measured before that change. Columns 1--4 relate debt composition in $t$ to this  value; column 5 imposes an additional exact one-year lag and therefore relates debt composition in $t$ to the  ETI dated $t-1$. Column 1 includes only the index and lagged emissions. Full models add lagged assets, debt/assets, tangibility, profitability, Tobin's Q and Altman Z-score. Columns 3 and 4 use exactly the same observations. Standard errors are clustered by firm and year; $p$-values use the number of year clusters minus one. Column 5 reports a one-standard-deviation effect from the native-coordinate exact-lag fit. The industry/year exact-lag estimate is $-0.00070$ (native-coordinate $p=.536$); because its two-way covariance is not positive semidefinite, alternative coordinate and unrepaired calculations give $p$-values from .374 to .627. The firm/year estimate shown in column 5 is stable to this issue. The common-sample contemporaneous estimate falls by 81.5\% when firm effects replace industry effects.
\end{tablenotes}
\end{threeparttable}
\end{table}

\begin{table}[!htbp]
\centering
 \caption{ETI and the composition of debt}
\label{tab:jfs_debt_components}
\begin{threeparttable}
\small
\begin{tabular}{@{}lccrrrr@{}}
\toprule
Outcome & Industry + year FE & Firm + year FE & $N_{IY}$ & $N_{FY}$ & Firms$_{IY}$ & Firms$_{FY}$ \\
\midrule
Bank debt share
 & \jfsresult{$-0.0914$}{0.0380}{.037}
 & \jfsresult{$-0.0178$}{0.0517}{.738} & 10,691 & 10,483 & 2,034 & 1,826 \\
Public bond debt share
 & \jfsresult{0.1234}{0.0736}{.125}
 & \jfsresult{0.0028}{0.0633}{.966} & 10,691 & 10,483 & 2,034 & 1,826 \\
Other debt share
 & \jfsresult{$-0.0320$}{0.0426}{.469}
 & \jfsresult{0.0150}{0.0427}{.733} & 10,691 & 10,483 & 2,034 & 1,826 \\
\bottomrule
\end{tabular}
\begin{tablenotes}[flushleft]\footnotesize
\item \textit{Notes:} All three outcomes are proportions of total debt and sum to one. Models use the  contemporaneous ETI, the full lagged control set, year effects and the fixed effects indicated. Standard errors are clustered by firm and year; unadjusted $p$-values use 10 degrees of freedom. Other debt is calculated as one minus the bank and bond shares. These share regressions do not identify public-bond issuance, green instruments, or the direction of borrower and lender decisions.
\end{tablenotes}
\end{threeparttable}
\end{table}

\begin{table}[!htbp]
\centering
 \caption{Inference and coefficient-sensitivity checks for the ETI model}
\label{tab:jfs_inference}
\label{tab:cluster_sensitivity}
\label{tab:loyo}
\label{tab:wild_year}
\begin{threeparttable}
\scriptsize
\begin{adjustbox}{max width=\textwidth}
\begin{tabular}{@{}llrrrrrr@{}}
\toprule
Panel & Specification & Coefficient & SE & $t$ & $p$ & $N$ & Firms \\
\midrule
\multicolumn{8}{@{}l}{\textit{Panel A: covariance estimators}}\\
 & Industry/year FE: homoskedastic OLS & $-0.09138$ & 0.06639 & $-1.376$ & .169 & 10,691 & 2,034 \\
 & Industry/year FE: firm clusters & $-0.09138$ & 0.06966 & $-1.312$ & .190 & 10,691 & 2,034 \\
 & Industry/year FE: year clusters & $-0.09138$ & 0.05513 & $-1.657$ & .128 & 10,691 & 2,034 \\
 & Industry/year FE: firm + year clusters & $-0.09138$ & 0.03797 & $-2.407$ & .037 & 10,691 & 2,034 \\
 & Firm/year FE: firm + year clusters & $-0.01779$ & 0.05172 & $-0.344$ & .738 & 10,483 & 1,826 \\
\addlinespace
\multicolumn{8}{@{}l}{\textit{Panel B: exact year-sign wild-cluster checks}}\\
 & Industry/year FE: year-studentized & $-0.09138$ & & $-1.657$ & .250 & 10,691 & 2,034 \\
 & Industry/year FE: resampled two-way statistic & $-0.09138$ & & $-2.407$ & .122 & 10,691 & 2,034 \\
 & Firm/year FE: year-studentized & $-0.01779$ & & $-0.337$ & .743 & 10,483 & 1,826 \\
 & Firm/year FE: resampled two-way statistic & $-0.01779$ & & $-0.344$ & .732 & 10,483 & 1,826 \\
\addlinespace
\multicolumn{8}{@{}l}{\textit{Panel C: leave-one-year-out summary}}\\
 & Industry/year FE: range across 11 omissions & [$-0.1251,-0.0575$] & & & & & \\
 & Industry/year FE: omissions with $p<.05$ & 6 of 11 & & & & & \\
 & Firm/year FE: range across 11 omissions & [$-0.0740,0.0108$] & & & & & \\
 & Firm/year FE: omissions with $p<.05$ & 0 of 11 & & & & & \\
\bottomrule
\end{tabular}
\end{adjustbox}
\begin{tablenotes}[flushleft]\footnotesize
\item \textit{Notes:} All rows use the full  model and its fixed-effect-specific estimation sample. The wild-cluster tests enumerate the 1,024 unique two-sided Rademacher sign assignments for 11 year clusters and impose the null. The year-studentized version is the preferred check with few year clusters; the resampled two-way statistic is also reported. In the industry/year model, the two-way variance is smaller than either one-way variance because the firm/year intersection component is subtracted. Leave-one-year-out $p$-values use the remaining ten year clusters. These checks concern inference and temporal influence; they do not establish exogeneity.
\end{tablenotes}
\end{threeparttable}
\end{table}

\begin{table}[!htbp]
\centering
 \caption{Loan spreads and the Emissions-to-Talk Index on unique DealScan facilities}
\label{tab:jfs_dealscan_spread}
\begin{threeparttable}
\scriptsize
\begin{adjustbox}{max width=\textwidth}
 \begin{tabular}{@{}lcccc@{}}
\toprule
 & (1) & (2) & (3) & (4) \\
\midrule
Fixed effects & Industry/year & Firm/year & Industry/year & Firm/year \\
Index timing & $t$ & $t$ & $t-1$ & $t-1$ \\
Emissions-to-Talk Index, one SD
 & \jfsresult{0.04735}{0.02674}{.115}
 & \jfsresult{0.00387}{0.02558}{.883}
 & \jfsresult{0.00538}{0.01618}{.749}
 & \jfsresult{$-0.03043$}{0.01985}{.169} \\
Facility and firm controls & Full & Full & Full & Full \\
Common sample within timing & Yes & Yes & Yes & Yes \\
Observations & 2,507 & 2,507 & 1,988 & 1,988 \\
Firms & 525 & 525 & 433 & 433 \\
Year clusters & 9 & 9 & 8 & 8 \\
\bottomrule
\end{tabular}
\end{adjustbox}
\begin{tablenotes}[flushleft]\footnotesize
\item \textit{Notes:} The outcome is the log all-in-drawn spread on accepted facilities  in the one-record-per-facility file, restricted to complete years 2010--2019. The Emissions-to-Talk Index is standardized over the master firm--years; columns~1--2 use the contemporaneous index and columns~3--4 an exact one-calendar-year lag. Facility controls are log amount, log maturity and security; firm controls are log assets, debt/assets, profitability, Altman Z-score and Tobin's Q. Unknown security status is treated as missing. The sample is held fixed across the two fixed-effect designs within each timing, and  standard errors are clustered by firm and year, with $p$-values based on the number of year clusters minus one. Both exact-lag covariance matrices  required a positive-semidefinite repair; neither contemporaneous covariance did.  The contemporaneous coefficient is smaller with firm effects, but none of the four estimates is statistically distinguishable from zero. Accepted-facility  spreads cannot separate loan demand from loan supply.
\end{tablenotes}
\end{threeparttable}
\end{table}

\clearpage
\section*{Figures}
\begin{figure}[!htbp]
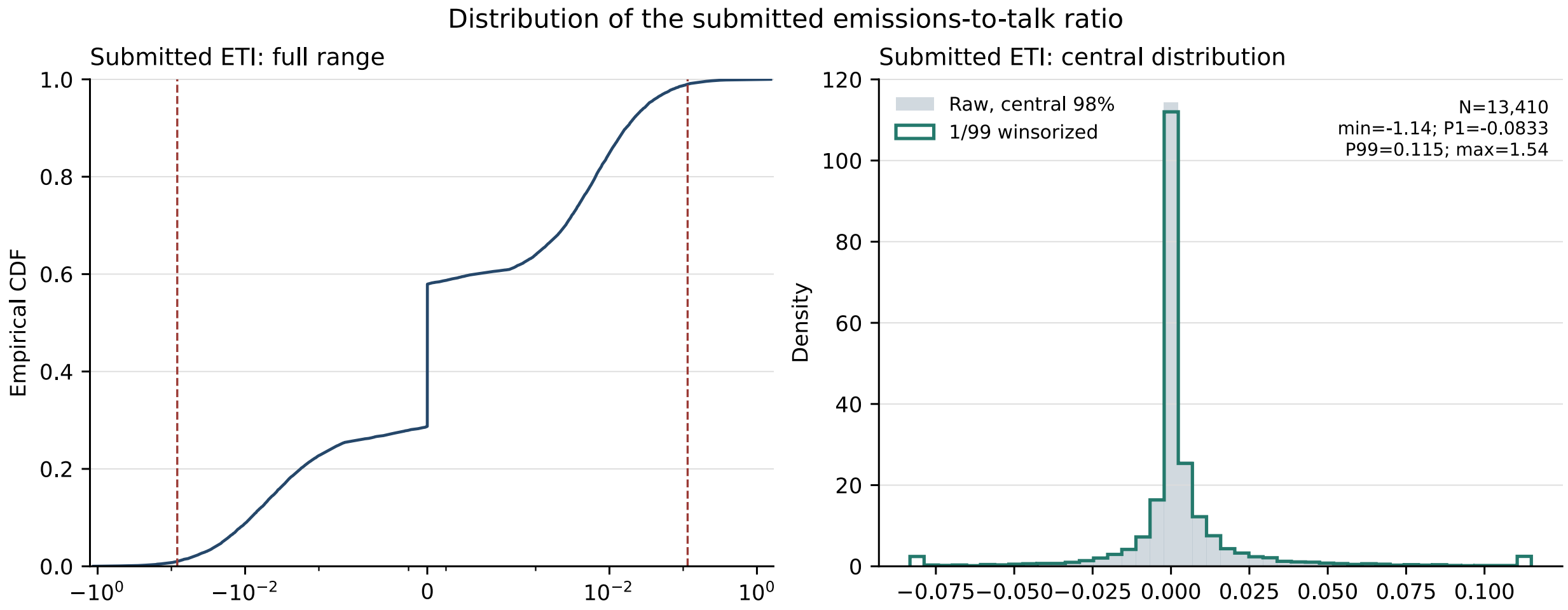

  \centering
  \resizebox{\textwidth}{!}{\JFSdistributionDrawing}
   \caption{Distribution of the Emissions-to-Talk Index}
  \label{fig:submitted_eti_distribution}
  \begin{minipage}{0.96\textwidth}
  \footnotesize
  \textit{Notes:} Panel A plots the empirical cumulative distribution over all
  13,410 nonmissing observations on a symmetric-log horizontal scale; the
  dashed lines mark the 1st and 99th percentiles. Panel B displays the central
  98 percent of the raw distribution and the corresponding fixed 1st/99th
  percentile winsorized distribution. The figure is descriptive.
  \end{minipage}
\end{figure}

\clearpage

\begin{figure}[!htbp]
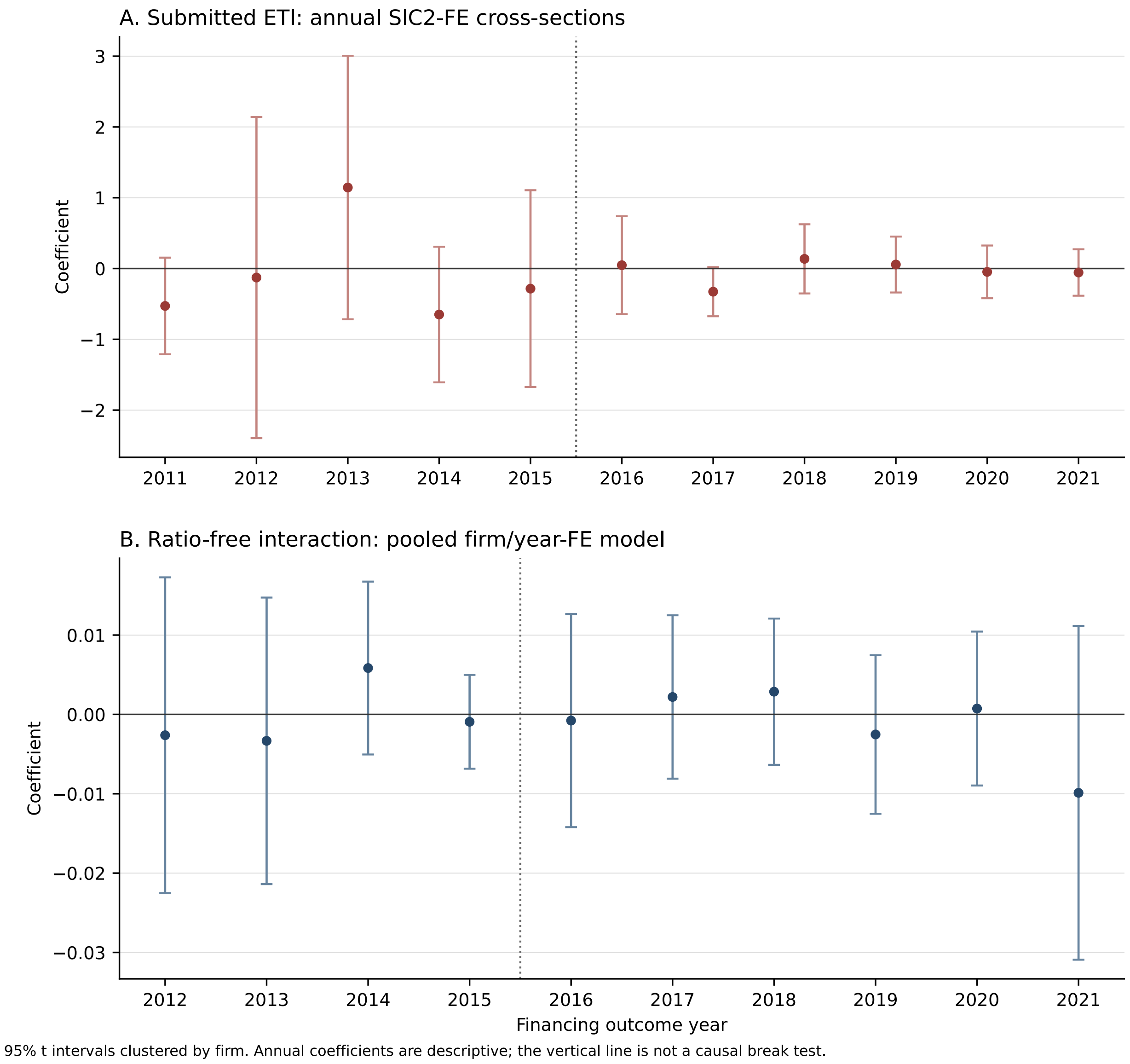

  \centering
  \adjustbox{max totalheight=0.78\textheight}{\resizebox{\textwidth}{!}{\JFSannualDrawing}}
  \caption{Year-specific debt-composition estimates}
  \label{fig:year_specific_combined}
  \begin{minipage}{0.96\textwidth}
  \footnotesize
  \textit{Notes:} Panel A reports year-specific industry-effect coefficients on the
  contemporaneous  ETI with two-digit-industry fixed effects.
  Panel B reports descriptive annual slopes for the exact-lag ratio-free
  interaction in the pooled firm/year specification. Bars are 95 percent
  confidence intervals. The dotted line separates pre-2016 and later outcome
  years for visual reference; it is not an event-study estimate or a causal
  break test.
  \end{minipage}
\end{figure}

\clearpage
\appendix
\setcounter{table}{0}
\renewcommand{\thetable}{A\arabic{table}}

\providecommand{\jfsresult}[3]{\shortstack{#1\\{\scriptsize (#2)}\\[-0.35ex]{\scriptsize $p=#3$}}}
\providecommand{\jfsdash}{\multicolumn{1}{c}{---}}

\begin{landscape}
\section{Supplementary tables and figures}\label{app:supplementary}
\begin{table}[!htbp]
\centering
 \caption{Distribution of the ETI and alternative ratio measures}
\label{tab:jfs_ratio_distribution}
\label{tab:ratio_distribution}
\begin{threeparttable}
\scriptsize
\begin{adjustbox}{max width=.95\linewidth}
 \begin{tabular}{@{}lrrrrrrrrrrr@{}}
\toprule
Measure & $N$ & Missing (\%) & Zero (\%) & Negative (\%) & Mean & SD & P1 & Median & P99 & Minimum & Maximum \\
\midrule
Emissions-to-Talk Index
 & 13,410 & 17.5 & 29.3 & 28.6 & 0.0030 & 0.0439 & $-0.0833$ & 0.0000 & 0.1149 & $-1.1368$ & 1.5409 \\
Presentation-only emissions/talk ratio
 & 9,519 & 41.5 & 29.6 & 29.2 & 168.7030 & 1,604.4850 & $-2,245.6787$ & 0.0000 & 4,311.1770 & $-8,117.3020$ & 69,705.1100 \\
Change in emissions/change in talk
 & 11,394 & 29.9 & 29.3 & 35.3 & $-3,606.8777$ & 382,359.4000 & $-12,390.2917$ & 0.0000 & 11,214.2303 & $-40,242,212.0000$ & 2,623,801.5000 \\
Presentation talk frequency
 & 16,040 & 1.3 & 29.0 & 0.0 & 0.001047 & 0.002714 & 0.000000 & 0.000261 & 0.013850 & 0.000000 & 0.046609 \\
\bottomrule
\end{tabular}
\end{adjustbox}
\begin{minipage}{.95\linewidth}\footnotesize
\textit{Notes:} Missing shares use all 16,258 firm--years. Zero shares are calculated among nonmissing observations. The presentation-only and change/change rows summarize the  additional ratio fields; they are not restricted to a regression sample. The former divides an emissions change by presentation talk, while the latter divides an emissions change by a change in presentation talk. The  ETI is also shown, in its raw units. The extreme ratio distributions motivate reporting results across definitions and sample rules.
\end{minipage}
\end{threeparttable}
\end{table}
\end{landscape}

\begin{table}[!htbp]
\centering
\caption{Pairwise correlations among principal climate-information measures}
\label{tab:jfs_correlations}
\label{tab:correlations}
\begin{threeparttable}
\scriptsize
\begin{adjustbox}{max width=\textwidth}
\begin{tabular}{@{}lrrrrrrrr@{}}
\toprule
 & (1) & (2) & (3) & (4) & (5) & (6) & (7) & (8) \\
\midrule
(1) Presentation climate talk       & 1.000 & $-0.047$ & 0.184 & 0.054 & 0.833 & 0.151 & 0.220 & $-0.037$ \\
(2) Change in log emissions         & $-0.017$ & 1.000 & $-0.075$ & $-0.042$ & $-0.054$ & $-0.089$ & $-0.102$ & 0.012 \\
(3) RepRisk environmental incidents & 0.100 & $-0.026$ & 1.000 & 0.402 & 0.182 & 0.358 & 0.405 & $-0.226$ \\
(4) RepRisk greenwashing incidents  & 0.055 & $-0.015$ & 0.724 & 1.000 & 0.050 & 0.222 & 0.218 & $-0.132$ \\
(5) Sautner overall climate exposure& 0.966 & $-0.018$ & 0.104 & 0.059 & 1.000 & 0.142 & 0.220 & $-0.029$ \\
(6) LSEG ESG score                  & 0.098 & $-0.047$ & 0.239 & 0.183 & 0.106 & 1.000 & 0.795 & $-0.220$ \\
(7) LSEG Emissions score            & 0.141 & $-0.058$ & 0.236 & 0.167 & 0.149 & 0.833 & 1.000 & $-0.226$ \\
(8) Bank debt share                 & $-0.070$ & 0.006 & $-0.131$ & $-0.096$ & $-0.068$ & $-0.265$ & $-0.279$ & 1.000 \\
\bottomrule
\end{tabular}
\end{adjustbox}
\begin{tablenotes}[flushleft]\footnotesize
\item \textit{Notes:} Pearson correlations are below the diagonal and Spearman rank correlations are above it. Pairwise $N$ ranges from 13,180 to 16,258. The corresponding source file contains exact $p$-values and pairwise observation counts. The 0.966 Pearson correlation between presentation talk and Sautner overall exposure reflects their closely related climate dictionaries, so the correlation does not provide an independent comparison of the two measures.
\end{tablenotes}
\end{threeparttable}
\end{table}

\begin{table}[!htbp]
\centering
 \caption{Exact-lag ETI and debt shares}
\label{tab:jfs_exact_lag_components}
\begin{threeparttable}
\small
\begin{tabular}{@{}lccrrr@{}}
\toprule
Outcome & Industry + year FE & Firm + year FE & $N$ & Firms & Years \\
\midrule
Bank debt share
 & \jfsresult{$-0.00070$}{0.00108}{.536}
 & \jfsresult{0.00159}{0.00297}{.606} & 8,334 & 1,560 & 10 \\
Public bond debt share
 & \jfsresult{0.00020}{0.00305}{.950}
 & \jfsresult{$-0.00024$}{0.00322}{.943} & 8,334 & 1,560 & 10 \\
Other debt share
 & \jfsresult{0.00050}{0.00235}{.837}
 & \jfsresult{$-0.00135$}{0.00234}{.578} & 8,334 & 1,560 & 10 \\
\bottomrule
\end{tabular}
\begin{tablenotes}[flushleft]\footnotesize
\item \textit{Notes:} The raw  ETI is lagged by exactly one calendar year; coefficients and standard errors from the native-coordinate regressions are then multiplied by the estimation-sample SD, 0.042737. Both fixed-effect specifications use the same observations, rebuilt lagged controls and two-way firm/year clustering. $p$-values use nine degrees of freedom. The industry/year covariance for the bank and other-debt models is not positive semidefinite and requires  an eigenvalue repair. For the bank-share model, alternative coordinate and unrepaired calculations give $p$-values from .374 to .627; .536 is the native-coordinate result. The firm/year results are stable. This table changes timing relative to the  contemporaneous specification and is a predictive-timing check rather than a direct coefficient comparison.
\end{tablenotes}
\end{threeparttable}
\end{table}

\begin{table}[!htbp]
\centering
\caption{Exact-lag index under additional common-shock fixed effects}
\label{tab:jfs_exact_lag_higher_fe}
\begin{threeparttable}
\small
\begin{tabular}{@{}lccrrr@{}}
\toprule
Fixed effects & One-SD estimate & Clustered SE & $p$ & $N$ & Firms \\
\midrule
\multicolumn{6}{@{}l}{\textit{Panel A: four-design common sample}}\\
Industry + year & $-0.000641$ & 0.001408 & .660 & 8,216 & 1,546 \\
Firm + year & 0.001844 & 0.003018 & .556 & 8,216 & 1,546 \\
Firm + two-digit SIC $\times$ year & 0.001512 & 0.002974 & .623 & 8,216 & 1,546 \\
Firm + state $\times$ year & 0.001477 & 0.002514 & .571 & 8,216 & 1,546 \\
\addlinespace
\multicolumn{6}{@{}l}{\textit{Panel B: each richer design's available sample}}\\
Firm + two-digit SIC $\times$ year & 0.001336 & 0.002954 & .662 & 8,277 & 1,554 \\
Firm + state $\times$ year & 0.001191 & 0.002599 & .657 & 8,278 & 1,552 \\
\bottomrule
\end{tabular}
\begin{tablenotes}[flushleft]\footnotesize
\item \textit{Notes:} The bank debt share is dated $t$; the  ETI and rebuilt controls are exact calendar lags. One-SD effects use the pooled exact-lag index standard deviation, 0.043628; this differs from the estimation-sample standard deviation used in Table~\ref{tab:jfs_exact_lag_components}. Standard errors are clustered by firm and year and use nine degrees of freedom. Panel A holds observations fixed across all four designs. State-by-year effects absorb state-specific common shocks; they are not a political instrument and do not supply a source of exogenous variation. None of the richer fixed-effect estimates is statistically distinguishable from zero.
\end{tablenotes}
\end{threeparttable}
\end{table}

\begin{table}[!htbp]
\centering
\caption{Alternative talk--emissions constructions and bank debt share}
\label{tab:jfs_alternative_measures}
\label{tab:alternative_measures}
\label{tab:ratio_debt}
\begin{threeparttable}
\scriptsize
\begin{adjustbox}{max width=\textwidth}
 \begin{tabular}{@{}lccrrrr@{}}
\toprule
Measure & Industry + year FE & Firm + year FE & $N_{IY}$ & $N_{FY}$ & Firms$_{IY}$ & Firms$_{FY}$ \\
\midrule
ETI
 & \jfsresult{$-0.09138$}{0.03797}{.037}
 & \jfsresult{$-0.01779$}{0.05172}{.738} & 10,691 & 10,483 & 2,034 & 1,826 \\
Presentation-only emissions/talk ratio
 & \jfsresult{$-3.37\times10^{-6}$}{$1.53\times10^{-6}$}{.053}
 & \jfsresult{$-4.54\times10^{-7}$}{$1.29\times10^{-6}$}{.731} & 7,822 & 7,482 & 1,851 & 1,511 \\
Change in emissions/change in talk
 & \jfsresult{$3.76\times10^{-9}$}{$1.25\times10^{-9}$}{.013}
 & \jfsresult{$2.71\times10^{-10}$}{$1.15\times10^{-9}$}{.818} & 9,248 & 9,007 & 1,925 & 1,684 \\
Standardized talk--emissions decoupling score
 & \jfsresult{0.00283}{0.00444}{.539}
 & \jfsresult{$-0.00210$}{0.00290}{.485} & 10,471 & 10,262 & 2,001 & 1,792 \\
\bottomrule
\end{tabular}
\end{adjustbox}
\begin{tablenotes}[flushleft]\footnotesize
\item \textit{Notes:} The dependent variable is the bank debt share. All rows use the full lagged control set and contemporaneous  measures, year effects, the indicated fixed effects, and two-way firm/year clustering. The ratio measures are reported in their  raw units, whereas the decoupling score is standardized. The change/change construction checks sensitivity to using changes in both components. Its denominator can be close to zero, and the sign of the ratio depends on whether talk rises or falls. The specifications therefore assess sensitivity to construction and should not be ranked by statistical significance.
\end{tablenotes}
\end{threeparttable}
\end{table}

\begin{landscape}
\begingroup
\renewcommand{\jfsresult}[3]{#1 (#2); $p=#3$}
\begin{table}[!htbp]
\centering
\caption{Exact-lag ratio tail checks and debt composition}
\label{tab:jfs_ratio_tail_diagnostics}
\begin{threeparttable}
\scriptsize
 \begin{tabular}{@{}llcccr@{}}
\toprule
Outcome & Alternative measure at $t-1$ & Firm + year FE & Industry + year FE & Holm $p_{FY}$ & $N$ \\
\midrule
Bank share & Change/change ratio & \jfsresult{$-0.0371$}{0.0357}{.325} & \jfsresult{$-0.1372$}{0.0720}{.089} & .325 & 7,197 \\
 & Change/change ratio, winsorized & \jfsresult{$-0.0017$}{0.0017}{.349} & \jfsresult{$-0.0030$}{0.0033}{.390} & 1.000 & 7,197 \\
 & Presentation-only ratio & \jfsresult{0.0013}{0.0020}{.518} & \jfsresult{$-0.0033$}{0.0025}{.221} & .717 & 5,929 \\
 & Presentation-only ratio, winsorized & \jfsresult{0.0081}{0.0022}{.005} & \jfsresult{0.0014}{0.0049}{.781} & .016 & 5,929 \\
 & ETI & \jfsresult{0.0016}{0.0030}{.606} & \jfsresult{$-0.0007$}{0.0014}{.626} & 1.000 & 8,334 \\
 & ETI, winsorized & \jfsresult{0.0069}{0.0031}{.054} & \jfsresult{0.0014}{0.0059}{.816} & .162 & 8,334 \\
\addlinespace
Bond share & Change/change ratio & \jfsresult{$-0.0685$}{0.0442}{.155} & \jfsresult{$-0.0778$}{0.0986}{.450} & .311 & 7,197 \\
 & Change/change ratio, winsorized & \jfsresult{0.0001}{0.0017}{.969} & \jfsresult{0.0014}{0.0037}{.716} & 1.000 & 7,197 \\
 & Presentation-only ratio & \jfsresult{0.0024}{0.0025}{.359} & \jfsresult{0.0011}{0.0017}{.543} & .717 & 5,929 \\
 & Presentation-only ratio, winsorized & \jfsresult{$-0.0009$}{0.0021}{.679} & \jfsresult{0.0075}{0.0042}{.109} & .679 & 5,929 \\
 & ETI & \jfsresult{$-0.0002$}{0.0033}{.943} & \jfsresult{0.0002}{0.0031}{.950} & 1.000 & 8,334 \\
 & ETI, winsorized & \jfsresult{$-0.0020$}{0.0038}{.602} & \jfsresult{0.0026}{0.0054}{.643} & .602 & 8,334 \\
\addlinespace
Other-debt share & Change/change ratio & \jfsresult{0.1056}{0.0532}{.078} & \jfsresult{0.2150}{0.0561}{.004} & .234 & 7,197 \\
 & Change/change ratio, winsorized & \jfsresult{0.0016}{0.0018}{.405} & \jfsresult{0.0016}{0.0017}{.367} & 1.000 & 7,197 \\
 & Presentation-only ratio & \jfsresult{$-0.0038$}{0.0017}{.052} & \jfsresult{0.0022}{0.0023}{.359} & .157 & 5,929 \\
 & Presentation-only ratio, winsorized & \jfsresult{$-0.0072$}{0.0026}{.021} & \jfsresult{$-0.0089$}{0.0027}{.009} & .042 & 5,929 \\
 & ETI & \jfsresult{$-0.0014$}{0.0024}{.578} & \jfsresult{0.0005}{0.0025}{.843} & 1.000 & 8,334 \\
 & ETI, winsorized & \jfsresult{$-0.0048$}{0.0029}{.133} & \jfsresult{$-0.0040$}{0.0042}{.362} & .265 & 8,334 \\
\bottomrule
\end{tabular}
\begin{tablenotes}[flushleft]\footnotesize
\item \textit{Notes:} Each ratio is standardized before its exact one-calendar-year lag is taken. The presentation-only and change/change rows use the rebuilt exact-calendar fields rather than the  alternative-measure fields summarized in Table~\ref{tab:jfs_ratio_distribution}. Winsorized variants apply fixed 1st/99th-percentile cutoffs before standardization. All specifications use common samples across fixed-effect designs and standard errors clustered by firm and year. Holm $p_{FY}$ adjusts the three debt-share outcomes within each firm-effect ratio specification. The industry/year  ETI models for bank and other-debt shares, and the firm/year untrimmed change/change model for other-debt share, required  a positive-semidefinite repair; no other model displayed here did. For the bank-share  ETI model, the standardized-coordinate result is $p=.626$; the native-coordinate and unrepaired alternatives are documented in Table~\ref{tab:jfs_exact_lag_components}, where the full range is $.374$--$.627$. These ratio estimates are checks on tail sensitivity, not preferred climate-information measures.
\end{tablenotes}
\end{threeparttable}
\end{table}
\endgroup
\end{landscape}

\begin{landscape}
\begin{table}[!htbp]
\centering
 \caption{Product-market competition: contemporaneous specification}
\label{tab:jfs_competition}
\label{tab:competition}
\begin{threeparttable}
\scriptsize
 \begin{tabular}{@{}lcccccc@{}}
\toprule
 & \multicolumn{2}{c}{Low substitutability} & \multicolumn{2}{c}{High concentration} & \multicolumn{2}{c}{Few competitors} \\
\cmidrule(lr){2-3}\cmidrule(lr){4-5}\cmidrule(l){6-7}
 & Industry/year & Firm/year & Industry/year & Firm/year & Industry/year & Firm/year \\
\midrule
ETI
 & \jfsresult{$-0.17996$}{0.08965}{.073} & \jfsresult{$-0.05799$}{0.07227}{.441}
 & \jfsresult{$-0.18305$}{0.07709}{.039} & \jfsresult{$-0.03673$}{0.06869}{.605}
 & \jfsresult{$-0.20107$}{0.07766}{.027} & \jfsresult{$-0.05790$}{0.06765}{.412} \\
Index $\times$ moderator
 & \jfsresult{0.30353}{0.14767}{.067} & \jfsresult{0.12041}{0.12225}{.348}
 & \jfsresult{0.33652}{0.09833}{.007} & \jfsresult{0.07375}{0.11174}{.524}
 & \jfsresult{0.41550}{0.11098}{.004} & \jfsresult{0.17021}{0.10121}{.124} \\
Moderator
 & \jfsresult{$-0.04480$}{0.01512}{.014} & \jfsresult{$-0.05343$}{0.01135}{.001}
 & \jfsresult{$-0.00554$}{0.00642}{.409} & \jfsresult{0.00738}{0.01252}{.569}
 & \jfsresult{$-0.01682$}{0.01806}{.374} & \jfsresult{0.00014}{0.01276}{.991} \\
Holm $p$, interaction family & .067 & .696 & .014 & .696 & .012 & .372 \\
Observations & 10,691 & 10,483 & 10,691 & 10,483 & 10,691 & 10,483 \\
$R^2$ & 0.2346 & 0.7844 & 0.2342 & 0.7837 & 0.2345 & 0.7838 \\
\bottomrule
\end{tabular}
\begin{tablenotes}[flushleft]\footnotesize
\item \textit{Notes:} The outcome is bank debt/total debt. These are  contemporaneous specifications using the ETI. Moderators are lagged one year. All specifications include the full  lagged control set and year effects; standard errors are clustered by firm and year. The Holm row adjusts the three interaction $p$-values separately within each fixed-effect design.  A four-firm concentration-ratio moderator  is not included among these splits. These interactions describe heterogeneity and do not identify a channel.
\end{tablenotes}
\end{threeparttable}
\end{table}
\end{landscape}

\begin{landscape}
\begin{table}[!htbp]
\centering
 \caption{Financial health: contemporaneous specification}
\label{tab:jfs_financial_health}
\label{tab:financial_health}
\begin{threeparttable}
\scriptsize
 \begin{tabular}{@{}lcccccc@{}}
\toprule
 & \multicolumn{2}{c}{Dividend payer} & \multicolumn{2}{c}{High Altman Z-score} & \multicolumn{2}{c}{High market capitalization} \\
\cmidrule(lr){2-3}\cmidrule(lr){4-5}\cmidrule(l){6-7}
 & Industry/year & Firm/year & Industry/year & Firm/year & Industry/year & Firm/year \\
\midrule
ETI
 & \jfsresult{$-0.13125$}{0.05429}{.036} & \jfsresult{$-0.03553$}{0.05884}{.559}
 & \jfsresult{$-0.20482$}{0.09502}{.057} & \jfsresult{$-0.06511$}{0.05638}{.275}
 & \jfsresult{$-0.10522$}{0.03438}{.012} & \jfsresult{$-0.01212$}{0.04982}{.813} \\
Index $\times$ moderator
 & \jfsresult{0.37167}{0.23016}{.137} & \jfsresult{0.17656}{0.14612}{.255}
 & \jfsresult{0.41406}{0.23877}{.114} & \jfsresult{0.20906}{0.10080}{.065}
 & \jfsresult{0.10575}{0.11191}{.367} & \jfsresult{$-0.02282$}{0.08549}{.795} \\
Moderator
 & \jfsresult{$-0.02381$}{0.01564}{.159} & \jfsresult{$-0.00403$}{0.01529}{.797}
 & \jfsresult{$-0.02432$}{0.01611}{.162} & \jfsresult{$-0.02489$}{0.00896}{.020}
 & \jfsresult{$-0.06714$}{0.01980}{.007} & \jfsresult{$-0.02187$}{0.01205}{.100} \\
Holm $p$, interaction family & .342 & .510 & .342 & .195 & .367 & .795 \\
Observations & 10,691 & 10,483 & 10,691 & 10,483 & 10,691 & 10,483 \\
$R^2$ & 0.2347 & 0.7838 & 0.2349 & 0.7840 & 0.2375 & 0.7839 \\
\bottomrule
\end{tabular}
\begin{tablenotes}[flushleft]\footnotesize
\item \textit{Notes:} Models and inference follow the contemporaneous  specification in Table~\ref{tab:jfs_competition}. The Holm row adjusts the three interaction $p$-values separately within each fixed-effect design. The high-Z-score interaction has raw $p=.065$ with firm effects and Holm $p=.195$.  A Whited--Wu-index interaction \citep{whitedwu2006}  is not included among these splits.
\end{tablenotes}
\end{threeparttable}
\end{table}
\end{landscape}

\begin{landscape}
\begin{table}[!htbp]
\centering
 \caption{Information opacity: contemporaneous specification}
\label{tab:jfs_opacity}
\label{tab:opacity}
\begin{threeparttable}
\scriptsize
 \begin{tabular}{@{}lcccccccc@{}}
\toprule
 & \multicolumn{2}{c}{High analyst dispersion} & \multicolumn{2}{c}{Low analyst coverage} & \multicolumn{2}{c}{High bid--ask proxy} & \multicolumn{2}{c}{High intangibles} \\
\cmidrule(lr){2-3}\cmidrule(lr){4-5}\cmidrule(lr){6-7}\cmidrule(l){8-9}
 & I/Y & F/Y & I/Y & F/Y & I/Y & F/Y & I/Y & F/Y \\
\midrule
ETI
 & \jfsresult{0.01402}{0.16219}{.933} & \jfsresult{0.08585}{0.10682}{.440}
 & \jfsresult{$-0.26307$}{0.09793}{.023} & \jfsresult{$-0.21251$}{0.07327}{.016}
 & \jfsresult{$-0.02742$}{0.10898}{.806} & \jfsresult{0.07990}{0.10609}{.469}
 & \jfsresult{$-0.09361$}{0.05000}{.091} & \jfsresult{$-0.00956$}{0.07350}{.899} \\
Index $\times$ moderator
 & \jfsresult{$-0.14943$}{0.23965}{.547} & \jfsresult{$-0.14891$}{0.12896}{.275}
 & \jfsresult{0.31324}{0.16416}{.085} & \jfsresult{0.34912}{0.13821}{.030}
 & \jfsresult{$-0.05738$}{0.12797}{.663} & \jfsresult{$-0.10796$}{0.09278}{.272}
 & \jfsresult{0.01732}{0.08896}{.850} & \jfsresult{$-0.01930$}{0.14567}{.897} \\
Moderator
 & \jfsresult{0.00087}{0.01033}{.935} & \jfsresult{$-0.00471$}{0.00745}{.541}
 & \jfsresult{0.07435}{0.01348}{<.001} & \jfsresult{0.00211}{0.01435}{.886}
 & \jfsresult{0.06925}{0.01267}{<.001} & \jfsresult{0.01854}{0.01051}{.108}
 & \jfsresult{0.06291}{0.01952}{.009} & \jfsresult{0.01206}{0.00993}{.252} \\
Holm $p$, interaction family & 1.000 & .816 & .340 & .120 & 1.000 & .816 & 1.000 & .897 \\
Observations & 10,691 & 10,483 & 10,691 & 10,483 & 10,691 & 10,483 & 10,691 & 10,483 \\
\bottomrule
\end{tabular}
\begin{tablenotes}[flushleft]\footnotesize
\item \textit{Notes:} I/Y and F/Y denote industry/year and firm/year fixed effects. These are  contemporaneous specifications using the ETI and binary opacity moderators; missing inputs are coded as low opacity. We therefore treat this table as a sensitivity check and use Table~\ref{tab:jfs_clean_exact_lag_interactions}, which leaves missing inputs missing, for the exact-lag comparison. All models include the full  lagged control set and use two-way firm/year clustering. The  Holm row adjusts the four interaction $p$-values within each fixed-effect design. The low-coverage interaction has raw $p=.030$ with firm effects and adjusted $p=.120$.  The source files do not document the construction of the bid--ask proxy. These four binary proxies do not cover every aspect of opacity.
\end{tablenotes}
\end{threeparttable}
\end{table}
\end{landscape}

\begin{landscape}
\begin{table}[!htbp]
\centering
\caption{Exact-lag interaction checks}
\label{tab:jfs_clean_exact_lag_interactions}
\begin{threeparttable}
\scriptsize
 \begin{tabular}{@{}llccrrr@{}}
\toprule
Family & Moderator at $t-1$ & Industry + year FE & Firm + year FE & $N$ & Firms & Years \\
\midrule
Competition & Low substitutability
 & \jfsresult{0.00769}{0.01819}{.682} & \jfsresult{$-0.00564$}{0.01422}{.701} & 8,334 & 1,560 & 10 \\
Competition & High concentration
 & \jfsresult{0.00335}{0.01586}{.838} & \jfsresult{$-0.00370$}{0.01753}{.837} & 8,334 & 1,560 & 10 \\
Competition & Few competitors
 & \jfsresult{0.00544}{0.01937}{.785} & \jfsresult{$-0.01078$}{0.01728}{.548} & 8,334 & 1,560 & 10 \\
\addlinespace
Financial health & Dividend payer
 & \jfsresult{0.02118}{0.01665}{.235} & \jfsresult{0.01638}{0.00760}{.060} & 8,305 & 1,558 & 10 \\
Financial health & High Altman Z-score
 & \jfsresult{0.01195}{0.02318}{.619} & \jfsresult{0.00209}{0.01755}{.908} & 8,334 & 1,560 & 10 \\
Financial health & High market capitalization
 & \jfsresult{$-0.01309$}{0.00944}{.199} & \jfsresult{$-0.01425$}{0.00604}{.043} & 8,334 & 1,560 & 10 \\
\addlinespace
Opacity & High analyst dispersion
 & \jfsresult{$-0.00059$}{0.00822}{.944} & \jfsresult{0.01137}{0.00835}{.206} & 6,568 & 1,261 & 10 \\
Opacity & Low analyst coverage
 & \jfsresult{$-0.00023$}{0.00618}{.971} & \jfsresult{0.00014}{0.00575}{.981} & 6,873 & 1,316 & 10 \\
Opacity & High bid--ask proxy
 & \jfsresult{0.01537}{0.00856}{.106} & \jfsresult{0.01311}{0.00586}{.052} & 8,276 & 1,555 & 10 \\
Opacity & High intangibles
 & \jfsresult{0.02000}{0.00889}{.051} & \jfsresult{0.01293}{0.00284}{.001} & 8,235 & 1,554 & 10 \\
\bottomrule
\end{tabular}
\begin{tablenotes}[flushleft]\footnotesize
\item \textit{Notes:} Entries are coefficients on the one-SD  ETI $\times$ moderator term, followed by standard errors and raw $p$-values. One-SD effects use the pooled exact-lag index standard deviation, 0.043628. The index, moderators and rebuilt controls are dated $t-1$ by exact calendar year; the debt-share outcome is dated $t$. Industry/year and firm/year models use the same observations within each row and standard errors are clustered by firm and year. Every industry/year covariance in this table required  a positive-semidefinite repair. Among the firm/year models, only high analyst dispersion and high intangibles required it. Holm adjustment within each family leaves no competition interaction below .10. In the financial-health family, the firm-effect dividend and market-capitalization interactions each have adjusted $p=.128$. In the opacity family, the firm-effect intangibles interaction has adjusted $p=.0055$; adjusting the two-way result across all ten interactions gives $p=.0137$. That result is coordinate-sensitive. An exact null-imposed year-cluster check gives $p=.043$ before and $.391$ after Holm adjustment across all ten moderators. The interaction is therefore exploratory rather than a general mechanism result.
\end{tablenotes}
\end{threeparttable}
\end{table}
\end{landscape}

\begin{table}[!htbp]
\centering
\caption{Pre/post-2016 estimates and coefficient-equality tests}
\label{tab:jfs_paris}
\label{tab:paris}
\begin{threeparttable}
\small
 \begin{tabular}{@{}lcc@{}}
\toprule
 & Industry + year FE & Firm + year FE \\
\midrule
\multicolumn{3}{@{}l}{\textit{Panel A: contemporaneous specification; split by outcome year}}\\
ETI $\times$ pre-2016
 & \jfsresult{$-0.00785$}{0.30553}{.980}
 & \jfsresult{$-0.05987$}{0.18470}{.753} \\
ETI $\times$ post-2016
 & \jfsresult{$-0.09489$}{0.03818}{.032}
 & \jfsresult{$-0.01683$}{0.05132}{.750} \\
Post minus pre
 & \jfsresult{$-0.08704$}{0.30744}{.783}
 & \jfsresult{0.04304}{0.17622}{.812} \\
Observations & 10,691 & 10,483 \\
Firms & 2,034 & 1,826 \\
$R^2$ & 0.2338 & 0.7837 \\
\addlinespace
\multicolumn{3}{@{}l}{\textit{Panel B: exact-lag model; split by outcome year}}\\
ETI $\times$ pre-2016
 & \jfsresult{0.00241}{0.01967}{.905}
 & \jfsresult{$-0.00432$}{0.01115}{.707} \\
ETI $\times$ post-2016
 & \jfsresult{$-0.00088$}{0.00142}{.552}
 & \jfsresult{0.00200}{0.00318}{.545} \\
Post minus pre
 & \jfsresult{$-0.00329$}{0.01958}{.870}
 & \jfsresult{0.00632}{0.01176}{.604} \\
Observations & 8,334 & 8,334 \\
Firms & 1,560 & 1,560 \\
\addlinespace
\multicolumn{3}{@{}l}{\textit{Panel C: exact-lag model; split by index source year}}\\
ETI $\times$ pre-2016
 & \jfsresult{0.00616}{0.01697}{.725}
 & \jfsresult{$-0.00401$}{0.00860}{.652} \\
ETI $\times$ post-2016
 & \jfsresult{$-0.00116$}{0.00141}{.432}
 & \jfsresult{0.00204}{0.00322}{.543} \\
Post minus pre
 & \jfsresult{$-0.00731$}{0.01683}{.674}
 & \jfsresult{0.00605}{0.00936}{.535} \\
Observations & 8,334 & 8,334 \\
Firms & 1,560 & 1,560 \\
\bottomrule
\end{tabular}
\begin{tablenotes}[flushleft]\footnotesize
\item \textit{Notes:} Panel A is the  contemporaneous specification using the ETI and fixed-effect-specific samples. Panels B and C use the one-SD index from exact $t-1$, with the pooled exact-lag standard deviation 0.043628, rebuilt controls from exact $t-1$, and a common industry/firm-effects sample; outcome years 2016--2021 are post in Panel B. Panel C dates the split by the index source year, so the 2016 outcome is assigned to the pre period because its regressor comes from 2015. Standard errors are clustered by firm and year. Each difference is a linear contrast of the two slopes using their covariance. Industry/year covariances in Panels B and C require  a positive-semidefinite repair; firm/year covariances do not. No equality test is below .50. Annual estimates for the contemporaneous specification appear in Figure~\ref{fig:year_specific_combined}, Panel A.
\end{tablenotes}
\end{threeparttable}
\end{table}

\begin{table}[!htbp]
\centering
\caption{Prior RepRisk incidents and bank debt share}
\label{tab:jfs_reprisk}
\label{tab:reprisk_specificity}
\begin{threeparttable}
\scriptsize
\begin{adjustbox}{max width=\textwidth}
\begin{tabular}{@{}llccrr@{}}
\toprule
Panel & Prior incident measure & Industry + year FE & Firm + year FE & $N_{IY}$ & $N_{FY}$ \\
\midrule
A & $\log(1+\text{all ESG incidents})$
 & \jfsresult{$-0.03665$}{0.00707}{<.001}
 & \jfsresult{$-0.01037$}{0.00476}{.054} & 10,691 & 10,483 \\
A & $\log(1+\text{greenwashing incidents})$
 & \jfsresult{$-0.02735$}{0.01822}{.164}
 & \jfsresult{0.00066}{0.00864}{.940} & 10,691 & 10,483 \\
\addlinespace
B & Environmental incidents, standardized
 & \jfsresult{$-0.02060$}{0.00701}{.015}
 & \jfsresult{$-0.00610$}{0.00228}{.023} & 10,373 & 10,373 \\
B & Any greenwashing incident
 & \jfsresult{0.01187}{0.01437}{.428}
 & \jfsresult{$-0.00156$}{0.00728}{.835} & 10,373 & 10,373 \\
B & All ESG incidents, standardized, standalone
 & \jfsresult{$-0.03844$}{0.00714}{<.001}
 & \jfsresult{$-0.01128$}{0.00475}{.039} & 10,373 & 10,373 \\
B & Environmental incidents conditional on all incidents
 & \jfsresult{0.00976}{0.00788}{.244}
 & \jfsresult{$-0.00156$}{0.00253}{.550} & 10,373 & 10,373 \\
\bottomrule
\end{tabular}
\end{adjustbox}
\begin{tablenotes}[flushleft]\footnotesize
\item \textit{Notes:} Panel A  uses logged-count specifications on each fixed-effect-specific sample. Panel B uses a common sample and distinguishes environmental incidents from a binary greenwashing indicator and total incident activity. Continuous Panel B measures are standardized. Models include the full lagged controls and two-way firm/year clustering. The environmental coefficient becomes small and insignificant when total incidents enter, while the greenwashing-specific coefficient is insignificant throughout. RepRisk records negative-news incidents; it is not another construction of the  emissions-to-talk index.
\end{tablenotes}
\end{threeparttable}
\end{table}

\begin{landscape}
\begin{table}[!htbp]
\centering
\caption{Coefficient-selection calculations and Frank ITCV}
\label{tab:jfs_oster_itcv}
\label{tab:eti_sensitivity}
\begin{threeparttable}
\scriptsize
 \begin{tabular}{@{}llrrrrrrr@{}}
\toprule
Fixed effects & Short model & $\hat\beta_S$ & $R^2_S$ & $\hat\beta_F$ & $R^2_F$ & $R_{\max}$ & Exact $\delta(\beta^*=0)$ & $N$ \\
\midrule
\multicolumn{9}{@{}l}{\textit{Panel A: proportional-selection calculations using within-transformed quantities}}\\
Industry/year & Index only & $-0.01980$ & 0.000007 & $-0.09138$ & 0.143822 & 0.186968 & $-4.267$ & 10,691 \\
Industry/year & Index only & $-0.01980$ & 0.000007 & $-0.09138$ & 0.143822 & 1.000000 & $-0.216$ & 10,691 \\
Industry/year & Index + lagged emissions & $-0.23283$ & 0.061301 & $-0.09138$ & 0.143822 & 0.186968 & 1.218 & 10,691 \\
Industry/year & Index + lagged emissions & $-0.23283$ & 0.061301 & $-0.09138$ & 0.143822 & 1.000000 & 0.0616 & 10,691 \\
Firm/year & Index only & $-0.02274$ & 0.000034 & $-0.01779$ & 0.008665 & 0.011265 & 5.812 & 10,483 \\
Firm/year & Index only & $-0.02274$ & 0.000034 & $-0.01779$ & 0.008665 & 1.000000 & 0.0153 & 10,483 \\
Firm/year & Index + lagged emissions & $-0.05116$ & 0.000329 & $-0.01779$ & 0.008665 & 0.011265 & 1.626 & 10,483 \\
Firm/year & Index + lagged emissions & $-0.05116$ & 0.000329 & $-0.01779$ & 0.008665 & 1.000000 & 0.0043 & 10,483 \\
\addlinespace
\multicolumn{9}{@{}l}{\textit{Panel B: Frank impact-threshold calculation for the industry/year estimate}}\\
\multicolumn{2}{l}{Rerun-result plug-in; conventional regression df $=10{,}608$} & \multicolumn{2}{l}{ITCV $=0.00442$} & \multicolumn{2}{l}{equal partial $r=0.06645$} & \multicolumn{3}{l}{reference $t=1.96019$} \\
\multicolumn{2}{l}{Cluster-calibrated analogue; year-cluster df $=10$} & \multicolumn{2}{l}{ITCV $=0.06986$} & \multicolumn{2}{l}{equal partial $r=0.26431$} & \multicolumn{3}{l}{reference $t=2.22814$} \\
\bottomrule
\end{tabular}
\begin{tablenotes}[flushleft]\footnotesize
\item \textit{Notes:} Panel A reports the exact coefficient-stability calculation with the full coefficient targeted at zero. Short and full models use the same observations and fixed effects. The full model adds lagged emissions and six financial controls to the index-only short model, or adds the six financial controls to the index-plus-emissions short model. $R_{\max}$ is either $\min(1,1.3R^2_F)$ or one. The large changes in $\delta$ across short-model and $R_{\max}$ choices show that the calculation is not a stand-alone identification result. Panel B applies Frank's published-results formula to the  specification's two-way-clustered $t$-statistic. The original ITCV derivation is for conventional OLS inference, so the cluster-calibrated row is an analogue, not an exact extension to multiway clustered covariance.
\end{tablenotes}
\end{threeparttable}
\end{table}
\end{landscape}

\begin{table}[!htbp]
\centering
 \caption{Within- and between-firm decomposition of the ETI}
\label{tab:jfs_within_between}
\begin{threeparttable}
\small
\begin{tabular}{@{}lccrr@{}}
\toprule
Component & Coefficient & Clustered SE & $p$ & $N$ \\
\midrule
Within-firm deviation & $-0.02557$ & 0.04467 & .580 & 10,483 \\
Firm mean & $-0.30679$ & 0.23543 & .222 & 10,483 \\
\bottomrule
\end{tabular}
\begin{tablenotes}[flushleft]\footnotesize
\item \textit{Notes:} Both components enter a model with industry and year effects and the  lagged controls. Standard errors are clustered by firm and year; $p$-values use ten degrees of freedom. The covariance estimate required  a positive-semidefinite safeguard. This decomposition is descriptive and does not replace the firm/year fixed-effect estimate.
\end{tablenotes}
\end{threeparttable}
\end{table}

\begin{landscape}
\begin{table}[!htbp]
\centering
 \caption{Firm--year DealScan outcomes and the Emissions-to-Talk Index}
\label{tab:jfs_dealscan_extensive}
\begin{threeparttable}
\scriptsize
 \begin{tabular}{@{}lcccccc@{}}
\toprule
 & \multicolumn{2}{c}{Any linked facility} & \multicolumn{2}{c}{Number of linked facilities} & \multicolumn{2}{c}{Log linked facility amount} \\
\cmidrule(lr){2-3}\cmidrule(lr){4-5}\cmidrule(l){6-7}
 & Industry/year & Firm/year & Industry/year & Firm/year & Industry/year & Firm/year \\
\midrule
Emissions-to-Talk Index, standardized & 0.008** & 0.000 & 0.017** & 0.002 & 0.173** & 0.005 \\
 & (0.003) & (0.003) & (0.007) & (0.006) & (0.062) & (0.058) \\
Lagged log emissions & 0.057*** & $-0.000$ & 0.109*** & 0.041 & 1.224*** & 0.044 \\
 & (0.010) & (0.019) & (0.020) & (0.027) & (0.209) & (0.388) \\
Full lagged controls & Yes & Yes & Yes & Yes & Yes & Yes \\
Observations & 10,912 & 10,714 & 10,912 & 10,714 & 10,912 & 10,714 \\
$R^2$ & 0.127 & 0.419 & 0.093 & 0.406 & 0.134 & 0.432 \\
\bottomrule
\end{tabular}
\begin{tablenotes}[flushleft]\footnotesize
\item \textit{Notes:}  These estimates are taken from the corrected workbook; the full aggregation path was not independently rerun for this table. The workbook describes firm--year aggregates built from the one-record-per-facility file: an indicator for any linked facility, the number of linked facilities, and log linked facility amount. The Emissions-to-Talk Index is standardized, and each model includes lagged log emissions and the full lagged control set. Industry/year and  firm/year models use fixed-effect-specific samples with two-way firm/year clustering.  Appearance in the linked extract measures matched borrowing, not all bank borrowing; a non-match does not establish that the firm obtained no bank credit elsewhere.  Ratio-free and RepRisk firm/year results appear in Table~\ref{tab:dealscan_firmyear}.
\end{tablenotes}
\end{threeparttable}
\end{table}
\end{landscape}

\begin{table}[!htbp]
\centering
 \caption{Non-price terms of accepted facilities and the Emissions-to-Talk Index}
\label{tab:jfs_dealscan_terms}
\begin{threeparttable}
\scriptsize
\begin{adjustbox}{max width=\textwidth}
\begin{tabular}{@{}lccrrrr@{}}
\toprule
Outcome & Industry + year FE & Firm + year FE & $N_{IY}$ & $N_{FY}$ & $R^2_{IY}$ & $R^2_{FY}$ \\
\midrule
Number of financial covenants
 & \jfsresult{$-0.034$}{0.019}{.085}
 & \jfsresult{0.018}{0.019}{.334} & 5,044 & 4,847 & 0.155 & 0.662 \\
Secured indicator
 & \jfsresult{0.005}{0.010}{.595}
 & \jfsresult{$-0.014$}{0.010}{.138} & 5,044 & 4,847 & 0.255 & 0.727 \\
Log maturity
 & \jfsresult{0.009}{0.008}{.225}
 & \jfsresult{0.007}{0.012}{.544} & 4,993 & 4,796 & 0.169 & 0.363 \\
\bottomrule
\end{tabular}
\end{adjustbox}
\begin{tablenotes}[flushleft]\footnotesize
\item \textit{Notes:}  These estimates are taken from the corrected workbook and were not independently rerun for this table. The outcomes are the number of financial covenants, a secured indicator and log maturity on accepted facilities in the one-record-per-facility file. The Emissions-to-Talk Index is standardized, and each model includes log facility amount and  the firm control set. Facilities with unknown security status  are coded as unsecured, whereas the analyses in Table~\ref{tab:dealscan_facility_main} leave them missing. Standard errors are clustered by firm. The displayed $p$-values were calculated from the workbook $t$-statistics using firm-cluster degrees of freedom and are shown to three decimals.  Ratio-free and RepRisk contract results with firm/year clustering appear in Table~\ref{tab:dealscan_facility_main}.
\end{tablenotes}
\end{threeparttable}
\end{table}

\begin{table}[!htbp]
\centering
\caption{Ratio-free climate-information terms and total debt}
\label{tab:jfs_total_debt}
\label{tab:total_debt}
\begin{threeparttable}
\scriptsize
\begin{adjustbox}{max width=\textwidth}
\begin{tabular}{@{}llcccrrr@{}}
\toprule
Outcome & Information term at $t-1$ & Firm + year FE & Industry + year FE & Holm $p_{FY}$ & $N$ & Firms & Years \\
\midrule
Log total debt & Talk $\times$ emissions change
 & \jfsresult{0.0023}{0.0049}{.651} & \jfsresult{0.0104}{0.0094}{.297} & 1.000 & 7,962 & 1,498 & 10 \\
Total debt/assets & Talk $\times$ emissions change
 & \jfsresult{$-0.0001$}{0.0010}{.916} & \jfsresult{0.0010}{0.0014}{.500} & 1.000 & 8,339 & 1,558 & 10 \\
Log total debt & High talk, rising emissions
 & \jfsresult{0.0314}{0.0222}{.190} & \jfsresult{0.0236}{0.0147}{.142} & .990 & 7,985 & 1,501 & 10 \\
Total debt/assets & High talk, rising emissions
 & \jfsresult{$-0.0043$}{0.0059}{.486} & \jfsresult{$-0.0022$}{0.0035}{.536} & 1.000 & 8,362 & 1,561 & 10 \\
Log total debt & Environmental incidents
 & \jfsresult{0.0038}{0.0095}{.696} & \jfsresult{0.0067}{0.0058}{.279} & .738 & 9,989 & 1,739 & 11 \\
Total debt/assets & Environmental incidents
 & \jfsresult{0.0047}{0.0021}{.054} & \jfsresult{0.0026}{0.0011}{.043} & .215 & 10,587 & 1,833 & 11 \\
Log total debt & Any greenwashing incident
 & \jfsresult{$-0.0242$}{0.0258}{.369} & \jfsresult{$-0.0294$}{0.0334}{.400} & .738 & 9,989 & 1,739 & 11 \\
Total debt/assets & Any greenwashing incident
 & \jfsresult{$-0.0078$}{0.0047}{.131} & \jfsresult{$-0.0070$}{0.0046}{.161} & .393 & 10,587 & 1,833 & 11 \\
\bottomrule
\end{tabular}
\end{adjustbox}
\begin{tablenotes}[flushleft]\footnotesize
\item \textit{Notes:} Total debt/assets is rebuilt as $(\texttt{dltt}+\texttt{dlc})/\texttt{assets}$; log total debt is defined when the sum is positive. Models include the lagged dependent variable, lagged emissions and the remaining lagged controls. Continuous information variables are standardized; the high-talk/rising and any-greenwashing indicators remain binary. Standard errors are clustered by firm and year. Holm $p_{FY}$ adjusts across both outcomes within each specification: six continuous coefficients, six quadrant indicators, or four RepRisk coefficients. The industry/year covariance required  a positive-semidefinite repair for both outcomes in the continuous and quadrant specifications and for total debt/assets, but not log total debt, in the RepRisk specification; no displayed firm/year covariance required it. The table addresses changes in the total-debt denominator without imputing bank or bond dollar amounts from shares.
\end{tablenotes}
\end{threeparttable}
\end{table}


\begin{table}[!htbp]
\centering
 \caption{Robustness of the ETI coefficient to added scores and absorbed effects}
\label{tab:jfs_archival_table13}
\label{tab:other_robustness}
\begin{threeparttable}
\scriptsize
\begin{adjustbox}{max width=\textwidth}
 \begin{tabular}{@{}lccrr@{}}
\toprule
Added control or absorbed effects & Industry-design estimate & Firm-design estimate & $N_{IY}$ & $N_{FY}$ \\
\midrule
Current LSEG ESG score
 & \jfsresult{$-0.09288$}{0.03884}{.038}
 & \jfsresult{$-0.01508$}{0.05037}{.771} & 10,690 & 10,482 \\
Current LSEG Emissions score
 & \jfsresult{$-0.09409$}{0.03718}{.030}
 & \jfsresult{$-0.01721$}{0.05163}{.746} & 10,688 & 10,481 \\
Lagged LSEG ESG score
 & \jfsresult{$-0.09588$}{0.03908}{.034}
 & \jfsresult{$-0.01683$}{0.05137}{.750} & 10,691 & 10,483 \\
Lagged LSEG Emissions score
 & \jfsresult{$-0.09537$}{0.03800}{.031}
 & \jfsresult{$-0.01764$}{0.05169}{.740} & 10,687 & 10,480 \\
Two-digit SIC $\times$ year effects
 & \jfsresult{$-0.08750$}{0.03088}{.018}
 & \jfsresult{$-0.03342$}{0.06043}{.592} & 10,637 & 10,429 \\
\bottomrule
\end{tabular}
\end{adjustbox}
\begin{tablenotes}[flushleft]\footnotesize
\item \textit{Notes:}  Estimates are for the ETI with the full lagged control set and firm/year clustering. ``Industry design'' means industry and year effects in the score rows and two-digit SIC-by-year effects in the final row. ``Firm design'' means firm and year effects in the score rows and firm plus two-digit-SIC-by-year effects in the final row.  None of the firm-design coefficients is statistically distinguishable from zero.
\end{tablenotes}
\end{threeparttable}
\end{table}

\providecommand{\jfsresult}[3]{\shortstack{#1\\{\scriptsize (#2)}\\[-0.35ex]{\scriptsize $p=#3$}}}

\begin{landscape}
\begin{table}[!htbp]
\centering
\caption{Ratio-free continuous climate-information terms and debt composition}
\label{tab:ratiofree_debt_main}
\begin{threeparttable}
\scriptsize
\begin{tabular}{@{}llcccrr@{}}
\toprule
Outcome & Information term at $t-1$ & Firm + year FE & Industry + year FE & Holm $p_{FY}$ & $N$ & Firms \\
\midrule
Bank share & Change in log emissions
 & \jfsresult{0.0059}{0.0031}{.092} & \jfsresult{0.0049}{0.0051}{.362} & .828 & 8,219 & 1,548 \\
Bank share & Presentation climate talk
 & \jfsresult{$-0.0028$}{0.0031}{.386} & \jfsresult{0.0062}{0.0053}{.270} & 1.000 & 8,219 & 1,548 \\
Bank share & Talk $\times$ emissions change
 & \jfsresult{$-0.0002$}{0.0011}{.865} & \jfsresult{0.0026}{0.0029}{.399} & 1.000 & 8,219 & 1,548 \\
Bond share & Change in log emissions
 & \jfsresult{$-0.0022$}{0.0047}{.646} & \jfsresult{$-0.0017$}{0.0059}{.786} & 1.000 & 8,219 & 1,548 \\
Bond share & Presentation climate talk
 & \jfsresult{0.0013}{0.0035}{.720} & \jfsresult{$-0.0067$}{0.0062}{.311} & 1.000 & 8,219 & 1,548 \\
Bond share & Talk $\times$ emissions change
 & \jfsresult{0.0024}{0.0014}{.129} & \jfsresult{$-0.0021$}{0.0033}{.531} & 1.000 & 8,219 & 1,548 \\
Other share & Change in log emissions
 & \jfsresult{$-0.0037$}{0.0029}{.231} & \jfsresult{$-0.0032$}{0.0029}{.289} & 1.000 & 8,219 & 1,548 \\
Other share & Presentation climate talk
 & \jfsresult{0.0016}{0.0026}{.568} & \jfsresult{0.0005}{0.0044}{.914} & 1.000 & 8,219 & 1,548 \\
Other share & Talk $\times$ emissions change
 & \jfsresult{$-0.0022$}{0.0016}{.192} & \jfsresult{$-0.0004$}{0.0024}{.856} & 1.000 & 8,219 & 1,548 \\
\bottomrule
\end{tabular}
\begin{tablenotes}[flushleft]\footnotesize
\item \textit{Notes:} Continuous terms are scaled by one standard deviation. All information terms and rebuilt controls use exact calendar lags; the outcome is dated $t$. The two fixed-effect designs use the same observations. Standard errors are clustered by firm and year and $p$-values use nine degrees of freedom. For bank share, the three firm-effect information terms are jointly insignificant ($p=.307$). Holm $p_{FY}$ adjusts the nine firm-effect information coefficients displayed in the table; the smallest adjusted value is .828.
\end{tablenotes}
\end{threeparttable}
\end{table}

\begin{table}[!htbp]
\centering
\caption{Exhaustively enumerated year-cluster wild-bootstrap checks}
\label{tab:jfs_wild_protocol}
\begin{threeparttable}
\scriptsize
\begin{adjustbox}{max width=\textwidth}
\begin{tabular}{@{}lrrrrrr@{}}
\toprule
Fixed-protocol coefficient & Years & Estimate & Year-CR1 $p$ & WCR $p$ & Two-way $p$ & Holm WCR $p$ \\
\midrule
Talk $\times$ emissions change $\rightarrow$ bank share & 10 & $-0.0002$ & .878 & .852 & .865 & .852 \\
High talk/rising emissions $\rightarrow$ bank share & 10 & 0.0101 & .144 & .117 & .193 & .352 \\
Environmental incidents $\rightarrow$ bank share & 11 & $-0.0061$ & .017 & .024 & .023 & .141 \\
High talk/rising emissions $\rightarrow$ any greenwashing incident & 10 & 0.0139 & .038 & .045 & .061 & .180 \\
Talk $\times$ emissions change $\rightarrow$ DealScan log spread & 8 & $-0.0090$ & .188 & .242 & .176 & .484 \\
High talk/rising emissions $\rightarrow$ DealScan facility count & 8 & 0.0335 & .027 & .023 & .043 & .141 \\
\bottomrule
\end{tabular}
\end{adjustbox}
\begin{tablenotes}[flushleft]\footnotesize
\item \textit{Notes:} The models use firm and year effects and the samples and controls specified for each regression. Year-CR1 clusters by outcome year only. WCR is the null-imposed, restricted-residual wild-cluster bootstrap-$t$ with every distinct two-sided Rademacher year-sign assignment enumerated: 128 assignments for eight years, 512 for ten, and 1,024 for eleven. The two-way column reports the firm/year-clustered result. Holm values adjust the six WCR tests displayed here. Exhaustive enumeration removes simulation error but does not make the bootstrap finite-sample exact under arbitrary errors. Because WCR clusters only by year, it does not model within-firm dependence and does not replace two-way clustering.
\end{tablenotes}
\end{threeparttable}
\end{table}
\end{landscape}

\begin{landscape}
\begin{table}[!htbp]
\centering
\caption{Debt composition across ratio-free talk--emissions configurations}
\label{tab:quadrant_debt}
\begin{threeparttable}
\scriptsize
\begin{tabular}{@{}llcccrr@{}}
\toprule
Outcome & Configuration at $t-1$ & Firm + year FE & Industry + year FE & Holm $p_{FY}$ & $N$ & Firms \\
\midrule
Bank share & High talk, non-rising emissions
 & \jfsresult{0.0064}{0.0053}{.256} & \jfsresult{$-0.0016$}{0.0118}{.895} & 1.000 & 8,242 & 1,551 \\
Bank share & High talk, rising emissions
 & \jfsresult{0.0101}{0.0072}{.193} & \jfsresult{$-0.0205$}{0.0155}{.217} & 1.000 & 8,242 & 1,551 \\
Bank share & Low talk, rising emissions
 & \jfsresult{0.0123}{0.0083}{.172} & \jfsresult{$-0.0258$}{0.0151}{.121} & 1.000 & 8,242 & 1,551 \\
Bond share & High talk, non-rising emissions
 & \jfsresult{$-0.0049$}{0.0056}{.403} & \jfsresult{0.0020}{0.0143}{.894} & 1.000 & 8,242 & 1,551 \\
Bond share & High talk, rising emissions
 & \jfsresult{0.0025}{0.0102}{.815} & \jfsresult{0.0333}{0.0148}{.051} & 1.000 & 8,242 & 1,551 \\
Bond share & Low talk, rising emissions
 & \jfsresult{0.0046}{0.0073}{.541} & \jfsresult{0.0451}{0.0136}{.009} & 1.000 & 8,242 & 1,551 \\
Other share & High talk, non-rising emissions
 & \jfsresult{$-0.0015$}{0.0053}{.785} & \jfsresult{$-0.0004$}{0.0060}{.952} & 1.000 & 8,242 & 1,551 \\
Other share & High talk, rising emissions
 & \jfsresult{$-0.0126$}{0.0067}{.094} & \jfsresult{$-0.0127$}{0.0088}{.184} & .755 & 8,242 & 1,551 \\
Other share & Low talk, rising emissions
 & \jfsresult{$-0.0169$}{0.0072}{.043} & \jfsresult{$-0.0193$}{0.0084}{.047} & .385 & 8,242 & 1,551 \\
\bottomrule
\end{tabular}
\begin{tablenotes}[flushleft]\footnotesize
\item \textit{Notes:} Low talk with non-rising emissions is the omitted configuration. Talk is split within two-digit SIC/year cells relative to the cell median among nonmissing presentation-talk values; cells require at least three observations. Rising emissions denotes a positive exact-calendar change. Information terms and rebuilt controls are dated $t-1$. Designs use a common sample and firm/year clustered standard errors. For bank share, the three firm-effect configuration indicators are jointly insignificant ($p=.407$). Holm $p_{FY}$ adjusts the nine firm-effect configuration coefficients; none is below .10.
\end{tablenotes}
\end{threeparttable}
\end{table}
\end{landscape}

\begin{landscape}
\begin{table}[!htbp]
\centering
\caption{Exploratory checks by earnings-call segment}
\label{tab:segment_checks}
\begin{threeparttable}
\scriptsize
\begin{tabular}{@{}lccrr@{}}
\toprule
Outcome at $t$ & Prepared presentation at $t-1$ & Answer section at $t-1$ & Holm $p$, prepared & Holm $p$, answer \\
\midrule
Change in log emissions
 & \jfsresult{0.003753}{0.005314}{.498} & \jfsresult{0.002660}{0.002981}{.395} & 1.000 & 1.000 \\
LSEG Emissions score
 & \jfsresult{$-0.072713$}{0.143263}{.623} & \jfsresult{0.224794}{0.149393}{.163} & 1.000 & 1.000 \\
LSEG ESG score
 & \jfsresult{$-0.055051$}{0.105003}{.612} & \jfsresult{0.066700}{0.091792}{.484} & 1.000 & 1.000 \\
Any RepRisk greenwashing incident
 & \jfsresult{0.007502}{0.003259}{.044} & \jfsresult{$-0.001904$}{0.001949}{.352} & .750 & 1.000 \\
$\log(1+\text{greenwashing incidents})$
 & \jfsresult{0.009568}{0.002367}{.002} & \jfsresult{$-0.001057$}{0.002366}{.665} & .049 & 1.000 \\
$\log(1+\text{environmental incidents})$
 & \jfsresult{0.005220}{0.005661}{.378} & \jfsresult{0.010714}{0.002978}{.005} & 1.000 & .097 \\
$\log(1+\text{misleading-communication incidents})$
 & \jfsresult{0.008396}{0.005213}{.138} & \jfsresult{0.001791}{0.004119}{.673} & 1.000 & 1.000 \\
\bottomrule
\end{tabular}
\begin{tablenotes}[flushleft]\footnotesize
\item \textit{Notes:} Each row is a joint firm/year-effects model containing standardized log-transformed presentation and answer frequencies, rebuilt lagged controls and the lagged dependent variable. Standard errors are clustered by firm and year. The row-specific sample sizes are 8,388 for emissions change; 10,508 for the Emissions score; 10,511 for the ESG score; and 10,512 for each incident outcome. These samples contain 1,565 and 1,827--1,828 firms, respectively. The last two columns are section-wide Holm adjustments over coefficient and joint tests.  Results are associations about incremental segment content, not causal effects.
\end{tablenotes}
\end{threeparttable}
\end{table}
\end{landscape}

\begin{table}[!htbp]
\centering
\caption{Ratio-free configurations and subsequent RepRisk incidents}
\label{tab:reprisk_checks}
\begin{threeparttable}
\scriptsize
\begin{adjustbox}{max width=\textwidth}
\begin{tabular}{@{}llccrr@{}}
\toprule
Outcome at $t$ & Configuration at $t-1$ & Firm + year FE & Industry + year FE & Holm $p_{FY}$ & $N$ \\
\midrule
Any greenwashing incident & High talk, rising emissions
 & \jfsresult{0.0139}{0.0065}{.061} & \jfsresult{0.0132}{0.0083}{.146} & .552 & 8,380 \\
Any greenwashing incident & Talk $\times$ emissions change
 & \jfsresult{0.0008}{0.0014}{.567} & \jfsresult{0.0004}{0.0014}{.762} & 1.000 & 8,357 \\
$\log(1+\text{environmental incidents})$ & High talk, rising emissions
 & \jfsresult{$-0.0005$}{0.0230}{.985} & \jfsresult{0.0654}{0.0300}{.057} & 1.000 & 8,380 \\
$\log(1+\text{environmental incidents})$ & Talk $\times$ emissions change
 & \jfsresult{0.0025}{0.0044}{.579} & \jfsresult{0.0035}{0.0064}{.597} & 1.000 & 8,357 \\
$\log(1+\text{greenwashing incidents})$ & High talk, rising emissions
 & \jfsresult{0.0113}{0.0054}{.066} & \jfsresult{0.0150}{0.0084}{.107} & .552 & 8,380 \\
$\log(1+\text{greenwashing incidents})$ & Talk $\times$ emissions change
 & \jfsresult{0.0011}{0.0011}{.371} & \jfsresult{0.0003}{0.0019}{.895} & 1.000 & 8,357 \\
\bottomrule
\end{tabular}
\end{adjustbox}
\begin{tablenotes}[flushleft]\footnotesize
\item \textit{Notes:} These regressions examine whether the measures predict later RepRisk incidents. Continuous terms are scaled by one SD, and all regressors use exact calendar lags. The binary-outcome specification is a linear probability model. Standard errors are clustered by firm and year. Holm values use the nine firm-effect coefficients in the relevant continuous or configuration family (three terms or configurations across three outcomes).
\end{tablenotes}
\end{threeparttable}
\end{table}

\begin{landscape}
\begin{table}[!htbp]
\centering
\caption{Joint regressions with climate-information measures}
\label{tab:external_horse_race}
\begin{threeparttable}
\scriptsize
\begin{tabular}{@{}lcccr@{}}
\toprule
Information at $t-1$ & Bank share & Bond share & Other share & Holm $p$ \\
\midrule
Presentation climate talk
 & \jfsresult{$-0.0024$}{0.0032}{.478} & \jfsresult{$-0.0001$}{0.0035}{.982} & \jfsresult{0.0024}{0.0027}{.383} & 1.000 \\
Change in log emissions
 & \jfsresult{0.0049}{0.0031}{.156} & \jfsresult{$-0.0012$}{0.0047}{.809} & \jfsresult{$-0.0037$}{0.0029}{.236} & 1.000 \\
Talk $\times$ emissions change
 & \jfsresult{$-0.0002$}{0.0010}{.829} & \jfsresult{0.0025}{0.0015}{.123} & \jfsresult{$-0.0022$}{0.0015}{.178} & 1.000 \\
RepRisk environmental incidents
 & \jfsresult{$-0.0023$}{0.0020}{.281} & \jfsresult{0.0009}{0.0035}{.797} & \jfsresult{0.0014}{0.0037}{.719} & 1.000 \\
Any RepRisk greenwashing incident
 & \jfsresult{0.0029}{0.0058}{.637} & \jfsresult{0.0023}{0.0071}{.754} & \jfsresult{$-0.0052$}{0.0093}{.593} & 1.000 \\
Sautner opportunity exposure
 & \jfsresult{$-0.0046$}{0.0069}{.522} & \jfsresult{0.0104}{0.0060}{.118} & \jfsresult{$-0.0058$}{0.0039}{.176} & 1.000 \\
Sautner regulatory exposure
 & \jfsresult{0.0016}{0.0020}{.432} & \jfsresult{$-0.0004$}{0.0019}{.840} & \jfsresult{$-0.0012$}{0.0027}{.658} & 1.000 \\
Sautner physical exposure
 & \jfsresult{0.0100}{0.0041}{.039} & \jfsresult{$-0.0055$}{0.0029}{.088} & \jfsresult{$-0.0045$}{0.0026}{.120} & 1.000 \\
LSEG ESG score
 & \jfsresult{$-0.0043$}{0.0102}{.680} & \jfsresult{0.0088}{0.0162}{.600} & \jfsresult{$-0.0045$}{0.0115}{.707} & 1.000 \\
LSEG Emissions score
 & \jfsresult{$-0.0031$}{0.0106}{.776} & \jfsresult{0.0077}{0.0119}{.535} & \jfsresult{$-0.0046$}{0.0081}{.587} & 1.000 \\
\midrule
Observations & \multicolumn{3}{c}{8,061} & \\
Firms & \multicolumn{3}{c}{1,529} & \\
\bottomrule
\end{tabular}
\begin{tablenotes}[flushleft]\footnotesize
\item \textit{Notes:} All ten information measures enter simultaneously in each firm/year-effects regression. Continuous measures are standardized; any greenwashing is binary. Rebuilt controls and information measures use exact calendar lags. Standard errors are clustered by firm and year. Holm $p$ adjusts the 30 displayed information coefficients; every adjusted value equals one. The firm/year bank-share covariance required  a positive-semidefinite repair; the bond- and other-share covariances did not.
\end{tablenotes}
\end{threeparttable}
\end{table}
\end{landscape}

\begin{table}[!htbp]
\centering
 \caption{Pairwise Pearson correlations: ETI and principal variables}
\label{tab:submitted_correlations}
\begin{threeparttable}
\scriptsize
\begin{adjustbox}{max width=\textwidth}
 \begin{tabular}{@{}lccccccc@{}}
\toprule
 & (1) & (2) & (3) & (4) & (5) & (6) & (7) \\
\midrule
(1) ETI & \shortstack{1.000\\[-1pt]{\tiny [13,410]}} & & & & & & \\
(2) Log total emissions
 & \shortstack{$-0.029^{***}$\\[-1pt]{\tiny [13,410]}} & \shortstack{1.000\\[-1pt]{\tiny [16,258]}} & & & & & \\
(3) TalkFreq
 & \shortstack{$-0.028^{***}$\\[-1pt]{\tiny [13,410]}} & \shortstack{$0.204^{***}$\\[-1pt]{\tiny [16,258]}} & \shortstack{1.000\\[-1pt]{\tiny [16,258]}} & & & & \\
(4) LSEG ESG score
 & \shortstack{$-0.041^{***}$\\[-1pt]{\tiny [13,409]}} & \shortstack{$0.411^{***}$\\[-1pt]{\tiny [16,257]}} & \shortstack{$0.104^{***}$\\[-1pt]{\tiny [16,257]}} & \shortstack{1.000\\[-1pt]{\tiny [16,257]}} & & & \\
(5) LSEG Emissions score
 & \shortstack{$-0.049^{***}$\\[-1pt]{\tiny [13,405]}} & \shortstack{$0.462^{***}$\\[-1pt]{\tiny [16,249]}} & \shortstack{$0.151^{***}$\\[-1pt]{\tiny [16,249]}} & \shortstack{$0.833^{***}$\\[-1pt]{\tiny [16,249]}} & \shortstack{1.000\\[-1pt]{\tiny [16,249]}} & & \\
(6) Firm size
 & \shortstack{$-0.037^{***}$\\[-1pt]{\tiny [13,409]}} & \shortstack{$0.529^{***}$\\[-1pt]{\tiny [16,257]}} & \shortstack{$0.097^{***}$\\[-1pt]{\tiny [16,257]}} & \shortstack{$0.567^{***}$\\[-1pt]{\tiny [16,256]}} & \shortstack{$0.582^{***}$\\[-1pt]{\tiny [16,248]}} & \shortstack{1.000\\[-1pt]{\tiny [16,257]}} & \\
(7) Bank debt share
 & \shortstack{0.004\\[-1pt]{\tiny [13,182]}} & \shortstack{$-0.272^{***}$\\[-1pt]{\tiny [15,912]}} & \shortstack{$-0.070^{***}$\\[-1pt]{\tiny [15,912]}} & \shortstack{$-0.265^{***}$\\[-1pt]{\tiny [15,911]}} & \shortstack{$-0.279^{***}$\\[-1pt]{\tiny [15,903]}} & \shortstack{$-0.351^{***}$\\[-1pt]{\tiny [15,911]}} & \shortstack{1.000\\[-1pt]{\tiny [15,912]}} \\
\bottomrule
\end{tabular}
\end{adjustbox}
\begin{tablenotes}[flushleft]\footnotesize
\item \textit{Notes:} Pairwise-complete observation counts appear in brackets. The  ETI is the emissions-growth-to-prior-talk ratio. TalkFreq combines the presentation and management-answer portions and is scaled by ten; scaling does not affect correlations. Variables are contemporaneous firm-year observations, although the  ETI formula embeds prior talk. Stars denote two-sided $p<.10$, $p<.05$ and $p<.01$.
\end{tablenotes}
\end{threeparttable}
\end{table}

\begin{landscape}
\begin{table}[!htbp]
\centering
\caption{RepRisk environmental incidents and bank debt share over time}
\label{tab:reprisk_dynamics}
\begin{threeparttable}
\scriptsize
\begin{adjustbox}{max width=\textwidth}
\begin{tabular}{@{}lllrrrrr@{}}
\toprule
Panel & Term & Fixed effects & Effect (pp) & 95\% CI (pp) & Raw $p$ & Holm $p$ & $N$ \\
\midrule
A & Incidents at $t-2$ & Industry/year & $-0.1209$ & [$-1.732$, 1.490] & .867 & 1.000 & 6,526 \\
A & Incidents at $t-1$ & Industry/year & $-0.5948$ & [$-2.468$, 1.278] & .485 & 1.000 & 6,526 \\
A & Incidents at $t$ & Industry/year & $-0.0174$ & [$-2.047$, 2.012] & .985 & 1.000 & 6,526 \\
A & Incidents at $t+1$ & Industry/year & $-1.0963$ & [$-2.786$, 0.593] & .173 & .692 & 6,526 \\
A & Incidents at $t-2$ & Firm/year & $-0.0567$ & [$-1.234$, 1.120] & .914 & 1.000 & 6,526 \\
A & Incidents at $t-1$ & Firm/year & $-0.4169$ & [$-1.151$, 0.317] & .227 & .906 & 6,526 \\
A & Incidents at $t$ & Firm/year & 0.1301 & [$-0.856$, 1.116] & .769 & 1.000 & 6,526 \\
A & Incidents at $t+1$ & Firm/year & $-0.6234$ & [$-2.009$, 0.762] & .330 & .990 & 6,526 \\
\addlinespace
B & Event time $\leq-3$ & Industry/year & $-10.9535$ & [$-17.386$, $-4.521$] & .004 & .011 & 7,636 \\
B & Event time $-2$ & Industry/year & $-6.6437$ & [$-12.831$, $-0.457$] & .038 & .038 & 7,636 \\
B & First incident year & Industry/year & $-10.2812$ & [$-17.242$, $-3.320$] & .008 & .016 & 7,636 \\
B & Event time $+1$ & Industry/year & $-12.3126$ & [$-16.965$, $-7.660$] & $<.001$ & $<.001$ & 7,636 \\
B & Event time $+2$ & Industry/year & $-14.6099$ & [$-20.312$, $-8.908$] & $<.001$ & $<.001$ & 7,636 \\
B & Event time $\geq+3$ & Industry/year & $-14.1559$ & [$-19.429$, $-8.883$] & $<.001$ & $<.001$ & 7,636 \\
B & Event time $\leq-3$ & Firm/year & $-2.6496$ & [$-7.553$, 2.253] & .256 & 1.000 & 7,636 \\
B & Event time $-2$ & Firm/year & 0.9370 & [$-1.102$, 2.975] & .330 & 1.000 & 7,636 \\
B & First incident year & Firm/year & $-0.4105$ & [$-2.193$, 1.372] & .619 & 1.000 & 7,636 \\
B & Event time $+1$ & Firm/year & $-0.5500$ & [$-2.941$, 1.841] & .619 & 1.000 & 7,636 \\
B & Event time $+2$ & Firm/year & $-3.0002$ & [$-6.751$, 0.751] & .105 & .630 & 7,636 \\
B & Event time $\geq+3$ & Firm/year & 1.2566 & [$-3.452$, 5.965] & .565 & 1.000 & 7,636 \\
\bottomrule
\end{tabular}
\end{adjustbox}
\begin{tablenotes}[flushleft]\footnotesize
\item \textit{Notes:} Panel A jointly includes the four displayed standardized incident terms and has nine year clusters. Panel B includes never-treated firms and firms whose first positive environmental incident appears in 2012--2019 after two observed zero-incident years; event time $-1$ is omitted and there are eleven year clusters. It is a descriptive two-way-fixed-effect event-time check, not a staggered-adoption causal design. Standard errors are clustered by firm and year. Both Panel B covariance matrices, for the industry/year and firm/year models, required  a positive-semidefinite repair; neither Panel A covariance did. The industry-effect pre-event differences and the firm-effect joint pre-event test ($p=.039$) caution against a causal reading.
\end{tablenotes}
\end{threeparttable}
\end{table}
\end{landscape}

\begin{landscape}
\begin{table}[!htbp]
\centering
\caption{Ratio-free climate information and terms of accepted DealScan facilities}
\label{tab:dealscan_facility_main}
\label{tab:dealscan_facility_industry}
\begin{threeparttable}
\scriptsize
\begin{tabular}{@{}llccccc@{}}
\toprule
Fixed effects & Information at $t-1$ & Log AISD spread & Log maturity & Secured & Covenants & Revolver \\
\midrule
Firm/year & Talk $\times$ emissions change
 & \jfsresult{$-0.0090$}{0.0060}{.176} & \jfsresult{0.0214}{0.0058}{.008} & \jfsresult{$-0.0046$}{0.0038}{.268} & \jfsresult{0.0117}{0.0166}{.504} & \jfsresult{$-0.0117$}{0.0044}{.032} \\
Firm/year & High talk, rising emissions
 & \jfsresult{$-0.0107$}{0.0296}{.729} & \jfsresult{0.0202}{0.0426}{.650} & \jfsresult{0.0010}{0.0194}{.962} & \jfsresult{$-0.0453$}{0.0558}{.444} & \jfsresult{$-0.0096$}{0.0326}{.777} \\
Firm/year & RepRisk environmental incidents
 & \jfsresult{0.0114}{0.0124}{.384} & \jfsresult{$-0.0022$}{0.0209}{.919} & \jfsresult{$-0.0082$}{0.0107}{.463} & \jfsresult{$-0.0335$}{0.0263}{.239} & \jfsresult{$-0.0028$}{0.0120}{.821} \\
Firm/year & Any RepRisk greenwashing incident
 & \jfsresult{$-0.0400$}{0.0473}{.422} & \jfsresult{$-0.0633$}{0.0970}{.532} & \jfsresult{$-0.0086$}{0.0280}{.766} & \jfsresult{$-0.0187$}{0.0677}{.789} & \jfsresult{0.0266}{0.0355}{.474} \\
\addlinespace
Industry/year & Talk $\times$ emissions change
 & \jfsresult{$-0.0042$}{0.0063}{.524} & \jfsresult{0.0118}{0.0066}{.120} & \jfsresult{$-0.0011$}{0.0045}{.811} & \jfsresult{0.0105}{0.0166}{.547} & \jfsresult{$-0.0009$}{0.0088}{.926} \\
Industry/year & High talk, rising emissions
 & \jfsresult{$-0.0363$}{0.0395}{.389} & \jfsresult{0.0221}{0.0436}{.627} & \jfsresult{0.0950}{0.0385}{.043} & \jfsresult{0.1191}{0.0660}{.114} & \jfsresult{$-0.0072$}{0.0255}{.786} \\
Industry/year & RepRisk environmental incidents
 & \jfsresult{$-0.0026$}{0.0169}{.883} & \jfsresult{$-0.0003$}{0.0171}{.987} & \jfsresult{0.0154}{0.0201}{.467} & \jfsresult{$-0.0316$}{0.0317}{.349} & \jfsresult{0.0171}{0.0056}{.015} \\
Industry/year & Any RepRisk greenwashing incident
 & \jfsresult{$-0.0326$}{0.0617}{.612} & \jfsresult{$-0.1505$}{0.0549}{.025} & \jfsresult{$-0.0711$}{0.0459}{.160} & \jfsresult{$-0.0862$}{0.1215}{.498} & \jfsresult{0.0599}{0.0341}{.117} \\
\bottomrule
\end{tabular}
\begin{tablenotes}[flushleft]\footnotesize
\item \textit{Notes:} The unique-facility file is restricted to complete years 2010--2019. Ratio-free models use 2012--2019 (eight year clusters); RepRisk models use 2011--2019 (nine). For log spread, interaction, configuration and RepRisk samples are 1,946/425 firms, 1,950/427 and 2,486/524. For the other outcomes they are 2,166/459, 2,172/460 and 2,755/560. Each industry/year row uses the same observations as its firm/year counterpart. Unknown security status is missing rather than coded as unsecured; this affects 5,518 facilities in the full file. Standard errors are clustered by firm and year. Recorded covariance repairs apply to: industry/year spread in both ratio-free specifications; maturity in both fixed-effect designs for both ratio-free specifications and in the firm/year RepRisk model; secured status in both fixed-effect designs for the continuous specification and in the firm/year quadrant model; firm/year covenants and revolver in the continuous specification; industry/year covenants and revolver in the quadrant specification; and industry/year revolver in the RepRisk specification. Each RepRisk model contains both displayed incident terms. No other model displayed here required repair. Within each fixed-effect design, the Holm families contain 15 continuous, 15 quadrant and 10 RepRisk tests. The smallest within-specification Holm $p$ is .115 and every global-primary Holm value equals one.
\end{tablenotes}
\end{threeparttable}
\end{table}
\end{landscape}

\begin{landscape}
\begin{table}[!htbp]
\centering
\caption{Climate information and firm--year outcomes in the supplied DealScan extract}
\label{tab:dealscan_firmyear}
\begin{threeparttable}
\scriptsize
\begin{tabular}{@{}lccccc@{}}
\toprule
Information at $t-1$ & Any facility & Log facility count & Log facility amount & Mean AISD spread & Mean maturity \\
\midrule
Talk $\times$ emissions change
 & \jfsresult{0.0034}{0.0048}{.498} & \jfsresult{$-0.0007$}{0.0047}{.878} & \jfsresult{0.0563}{0.0979}{.583} & \jfsresult{$-2.1090$}{1.2529}{.136} & \jfsresult{0.5907}{0.3362}{.122} \\
High talk, rising emissions
 & \jfsresult{0.0250}{0.0175}{.198} & \jfsresult{0.0335}{0.0136}{.043} & \jfsresult{0.5481}{0.3520}{.163} & \jfsresult{1.3377}{4.8651}{.791} & \jfsresult{0.7711}{0.8011}{.368} \\
RepRisk environmental incidents
 & \jfsresult{0.0048}{0.0127}{.714} & \jfsresult{0.0022}{0.0093}{.816} & \jfsresult{0.0835}{0.2718}{.767} & \jfsresult{0.3908}{1.9263}{.844} & \jfsresult{$-0.3566$}{0.5173}{.510} \\
Any RepRisk greenwashing incident
 & \jfsresult{$-0.0092$}{0.0276}{.747} & \jfsresult{0.0048}{0.0341}{.892} & \jfsresult{$-0.1942$}{0.6131}{.760} & \jfsresult{$-1.9342$}{6.2773}{.766} & \jfsresult{$-1.5878$}{1.7712}{.396} \\
\bottomrule
\end{tabular}
\begin{tablenotes}[flushleft]\footnotesize
\item \textit{Notes:} Firm and year effects and exact-lag controls are included. The first three outcomes measure appearance or volume in the supplied extract; a zero does not establish that the firm obtained no bank loan elsewhere. Interaction/configuration samples are 6,572/6,596 observations for the first three outcomes, 1,639/1,642 for mean spread and 1,913/1,918 for mean maturity. RepRisk samples are 8,645, 2,170 and 2,508, respectively. Standard errors are clustered by firm and year. Recorded covariance repairs apply to mean spread in all three information specifications and, in the quadrant specification, to log facility count and log facility amount. Each RepRisk model contains both displayed incident terms. No other displayed model required repair. The Holm families contain 15 continuous, 15 quadrant and 10 RepRisk tests. The smallest specification-level Holm value is .646.
\end{tablenotes}
\end{threeparttable}
\end{table}
\end{landscape}

\begin{landscape}
\begingroup
\renewcommand{\jfsresult}[3]{#1 (#2); $p=#3$}
\begin{table}[!htbp]
\centering
\caption{DealScan contract-composition and USD-only checks}
\label{tab:dealscan_contract_robustness}
\begin{threeparttable}
\scriptsize
\begin{adjustbox}{max width=\textwidth}
\begin{tabular}{@{}lllcrrr@{}}
\toprule
Sample & Outcome & Information at $t-1$ & Firm/year estimate & Holm $p$ & $N$ & Year clusters \\
\midrule
All currencies & Log spread & Talk $\times$ emissions change & \jfsresult{$-0.0104$}{0.0053}{.087} & 1.000 & 1,946 & 8 \\
All currencies & Log spread & High talk, rising emissions & \jfsresult{$-0.0113$}{0.0290}{.707} & 1.000 & 1,950 & 8 \\
All currencies & Log spread & RepRisk environmental incidents & \jfsresult{0.0085}{0.0155}{.598} & 1.000 & 2,486 & 9 \\
All currencies & Log spread & Any RepRisk greenwashing incident & \jfsresult{$-0.0091$}{0.0506}{.861} & 1.000 & 2,486 & 9 \\
All currencies & Secured & Talk $\times$ emissions change & \jfsresult{$-0.0040$}{0.0035}{.294} & 1.000 & 2,166 & 8 \\
All currencies & Secured & High talk, rising emissions & \jfsresult{0.0012}{0.0205}{.957} & 1.000 & 2,172 & 8 \\
All currencies & Secured & RepRisk environmental incidents & \jfsresult{$-0.0092$}{0.0110}{.425} & 1.000 & 2,755 & 9 \\
All currencies & Secured & Any RepRisk greenwashing incident & \jfsresult{$-0.0150$}{0.0249}{.564} & 1.000 & 2,755 & 9 \\
USD only & Log spread & Talk $\times$ emissions change & \jfsresult{$-0.0109$}{0.0052}{.075} & 1.000 & 1,854 & 8 \\
USD only & Log spread & High talk, rising emissions & \jfsresult{$-0.0015$}{0.0265}{.955} & 1.000 & 1,858 & 8 \\
USD only & Log spread & RepRisk environmental incidents & \jfsresult{0.0124}{0.0149}{.429} & 1.000 & 2,366 & 9 \\
USD only & Log spread & Any RepRisk greenwashing incident & \jfsresult{0.0081}{0.0501}{.875} & 1.000 & 2,366 & 9 \\
USD only & Secured & Talk $\times$ emissions change & \jfsresult{$-0.0041$}{0.0037}{.303} & 1.000 & 2,033 & 8 \\
USD only & Secured & High talk, rising emissions & \jfsresult{0.0023}{0.0186}{.907} & 1.000 & 2,039 & 8 \\
USD only & Secured & RepRisk environmental incidents & \jfsresult{$-0.0083$}{0.0115}{.492} & 1.000 & 2,588 & 9 \\
USD only & Secured & Any RepRisk greenwashing incident & \jfsresult{$-0.0172$}{0.0258}{.523} & 1.000 & 2,588 & 9 \\
\bottomrule
\end{tabular}
\end{adjustbox}
\begin{tablenotes}[flushleft]\footnotesize
\item \textit{Notes:} All-currency models add grouped loan-type, deal-purpose and currency indicators; USD-only models add grouped type and purpose indicators. Groups with fewer than 20 facilities are pooled. Controls and complete-year restrictions match Table~\ref{tab:dealscan_facility_main}; unknown security status is missing rather than coded as unsecured. Each of the 12 firm/year models represented here (two samples $\times$ two outcomes $\times$ three information specifications) required  a positive-semidefinite repair. All Holm-adjusted values equal one.
\end{tablenotes}
\end{threeparttable}
\end{table}
\endgroup
\end{landscape}

\clearpage
\setcounter{figure}{0}
\renewcommand{\thefigure}{A\arabic{figure}}
\begin{figure}[!htbp]
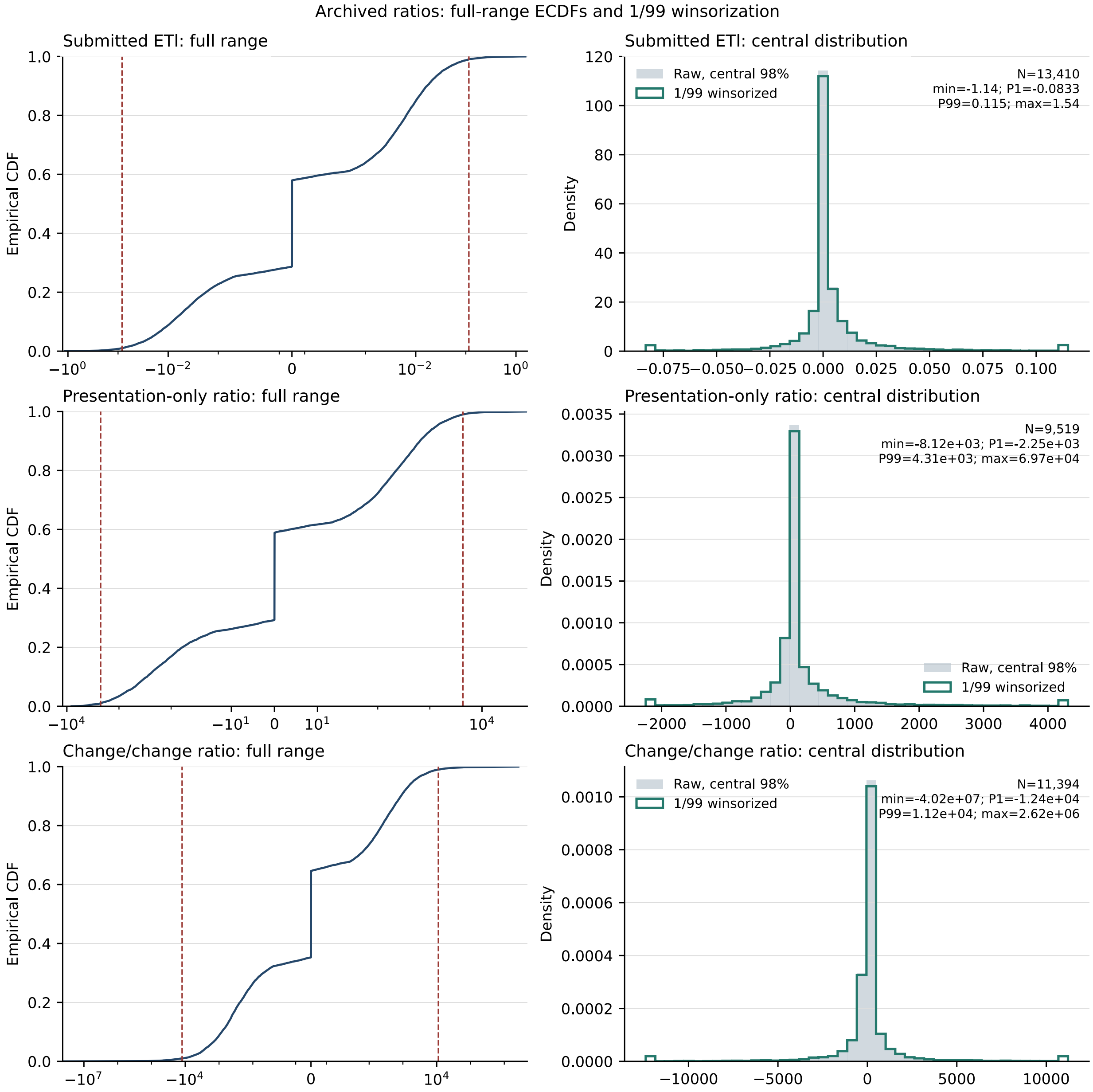

  \centering
  \adjustbox{max totalheight=0.78\textheight}{\resizebox{\textwidth}{!}{\JFSratiosDrawing}}
  \caption{Distributions of three emissions-to-talk ratios}
  \label{fig:all_ratio_distributions}
  \begin{minipage}{0.96\textwidth}
  \footnotesize
  \textit{Notes:} The rows show the  ETI, the  presentation-only
  emissions-growth-to-talk ratio, and the  emissions-change-to-talk-change
  ratio. Left panels plot empirical cumulative distributions on symmetric-log
  scales, with dashed lines at the 1st and 99th percentiles. Right panels show
  the central 98 percent of each raw distribution and its fixed 1st/99th
  percentile winsorized counterpart. The  three ratios are shown in their
  raw units.
  \end{minipage}
\end{figure}

\end{document}